\documentclass[useAMS,usenatbib]{mn2e}

\usepackage{graphicx}
\usepackage{subfig}

\title[Cepheids in Open Clusters]{Cepheids in Open Clusters: An 8-D All-sky
Census\thanks{Based on observations collected at the ESO La Silla Observatory
with the CORALIE echelle spectrograph mounted to the Swiss 1.2m Euler
telescope.}\thanks{Based on observations obtained with the HERMES spectrograph,
which is supported by the Fund for Scientific Research of Flanders (FWO),
Belgium , the Research Council of K.U. Leuven, Belgium, the Fonds National
Recherches Scientific (FNRS), Belgium, the Royal Observatory of Belgium, the
Observatoire de Gen\`eve, Switzerland and the Th\"uringer Landessternwarte
Tautenburg, Germany. HERMES is  mounted to the Mercator Telescope, operated on the island of
La Palma by the Flemish Community, at the Spanish Observatorio del Roque de los Muchachos of
the Instituto de Astrof\'isica de Canarias.}}
 
\author[R.I.~Anderson et al.]{Richard I. Anderson$^1$\thanks{E-mail:
\texttt{richard.anderson@unige.ch}}, Laurent Eyer$^1$, and Nami Mowlavi$^1$\\
$^1$ Observatoire de Gen\`eve, Universit\'e de Gen\`eve, 51 Ch. des Maillettes,
CH-1290 Versoix, Switzerland } 

\begin{document} 

\pagerange{\pageref{firstpage}--\pageref{lastpage}} \pubyear{2012}

\maketitle

\label{firstpage}
\begin{abstract}
Cepheids in open clusters (cluster Cepheids: CCs) are of great importance as
zero-point calibrators of the Galactic Cepheid period-luminosity relationship
(PLR).

We perform an 8-dimensional all-sky census that aims to identify new
\textit{bona-fide} CCs and provide a ranking of membership confidence for known
CC candidates according to membership probabilities. The probabilities
are computed for combinations of known Galactic open clusters and classical Cepheid
candidates, based on spatial, kinematic, and population-specific membership
constraints.
Data employed in this analysis are taken largely from published literature and
supplemented by a year-round observing program on both hemispheres dedicated to
determining systemic radial velocities of Cepheids.

In total, we find 23 \textit{bona-fide} CCs, 5 of which are
candidates identified for the first time, including an overtone-Cepheid member
in NGC\,129.
We discuss a subset of CC candidates in detail, some of
which have been previously mentioned in the literature.
Our results indicate unlikely membership for 7 Cepheids
that have been previously discussed in terms of cluster membership.

We furthermore revisit the Galactic PLR using our \textit{bona fide} CC sample
and obtain a result consistent with the recent calibration by
\cite{2010Ap&SS.326..219T}. However, our calibration remains limited mainly by
cluster uncertainties and the small number of long-period calibrators. 

In the near future, Gaia will enable our study to be carried out in much greater detail and accuracy, thanks to data homogeneity and greater levels of completeness.

\end{abstract}


\section{Introduction}\label{sec:Intro}
The search for Cepheids in Galactic open clusters (CCs) has been a topic of
interest in astronomy for the past $60$ years, owing largely to their importance
as calibrators of the Cepheid period-luminosity relation (PLR), discovered a
century ago among $25$ periodic variable stars in the SMC by
\cite{1912HarCi.173....1L}.

The proportionality between the logarithm of Cepheid pulsation periods and their
absolute magnitudes, i.e., their (logarithmic) luminosities, gives
access to distance determinations and has established period-luminosity relationships as
cornerstones of the astronomical distance scale
\citep[e.g.][]{2001ApJ...553...47F,2006ApJ...653..843S}. For reviews on Cepheids
as distance indicators, cf. \cite{1999PASP..111..775F,2006ARA&A..44...93S}, for
instance.

The existence of the Cepheid PLR is most obvious among Cepheids in the
Magellanic Clouds
\citep[e.g.][]{1999AcA....49..223U,2008AcA....58..163S,2010AcA....60...17S}, due
to common distances (small dispersion), large statistics (thousands),
and relative proximity (detectability). However, knowledge of the zero-point(s)
of such relations is also required; in this case, the distances to the
Magellanic Clouds. For such zero-point calibrations, PLR-independent distance
estimates are required, e.g. from trigonometric parallaxes
\citep{1997MNRAS.286L...1F,2007AJ....133.1810B}, Baade-Wesselink-type methods
\citep{1997ApJ...488...74G,2011A&A...534A..94S}, or objects located at
comparable distance, e.g. water masers \citep{2006ApJ...652.1133M} or open
clusters \citep{2010OAP....23..119T}.

For open clusters, distances can be determined via zero-age Main Sequence or
isochrone fitting. If membership can be assumed at high confidence, the cluster
provides the independent estimation of the Cepheid's distance. Confidence in
cluster membership is thus critical for such calibrations.

Since the first discovery of CCs by \citet[identified S\,Nor in NGC\,6087 and
U\,Sgr in M\,25]{1955MNSSA..14...38I} and \citet[established membership via
radial velocities]{1957MNRAS.117..193F}, many researchers have contributed to
this field, e.g.
\cite{1956PZ.....11..325K,1957ApJ...126..323V,1964PZ.....15..242E,1966ATsir.367....1T,
1986AJ.....92..111T, 1993ApJS...85..119T, 2000A&AS..146..251B,
2003MNRAS.345..269H, 2007ApJ...671.1640A, 2008MNRAS.390.1539M,
2010Ap&SS.326..219T}. Nevertheless, relatively few \textit{bona-fide} CCs ($<
30$) have thus far been discovered.

We therefore carry out an all-sky census of classical Cepheids in Galactic Open
Clusters that aims to increase the number of \textit{bona-fide} CCs and allows
us to rank confidence in membership according to membership probabilities. Our
approach is 8 dimensional in the sense that 3 spatial, 3 kinematic, and two
population parameters (iron abundance and age) are used as membership
constraints. Both data inhomogeneity and incompleteness are critical limitations
to this work, and are acknowledged in the relevant sections.
We describe our analysis in Sec.\,\ref{sec:memcalc}.

For the first time, we systematically search for cluster members among Cepheid
candidates from surveys such as ASAS, NSVS, ROTSE, and also from the suspected
variables in the General Catalog of Variable Stars.
Most data employed to do so are taken from published catalogs or other
literature.  However, we also perform radial velocity observations of Cepheids
on both hemispheres and determine systemic velocities, $v_\gamma$. To improve
sensitivity to binarity, literature RVs are added to the new observations.
The data compilation is described in Sec.\,\ref{sec:LitData}.

The results of our census are presented in Sec.\,\ref{sec:Results},
starting with cluster-Cepheid combinations (Combos) that were previously studied
with respect to membership, see Tab.\,\ref{tab:literature} in
Sec.\,\ref{sec:litCombos}, and followed by Combos highlighted by our work, see
Tab.\,\ref{tab:newCombos} in Sec.\,\ref{sec:newCombos}.
The full table containing all Combos investigated in this work is
provided in digital form in the online appendix and via the
CDS\footnote{\texttt{http://cds.u-strasbg.fr}}. An example of the information provided can be found in Tab.\,\ref{tab:onlinetable}. Combos that deserve
observational follow-up are identified in the text. Particular attention is
given to Combos previously discussed in the literature. Discussions of
additional Combos can be found in the online appendix. In
Sec.\,\ref{sec:GalPLR}, we employ our \textit{bona-fide} CC sample in a
calibration of the Galactic Cepheid PLR. The method and results are discussed in
Sec.\,\ref{sec:Discussion}, which is followed by the conclusion in
Sec.\,\ref{sec:conclusion}.

\section{Membership Analysis}\label{sec:Analysis} 

\begin{figure}
\includegraphics{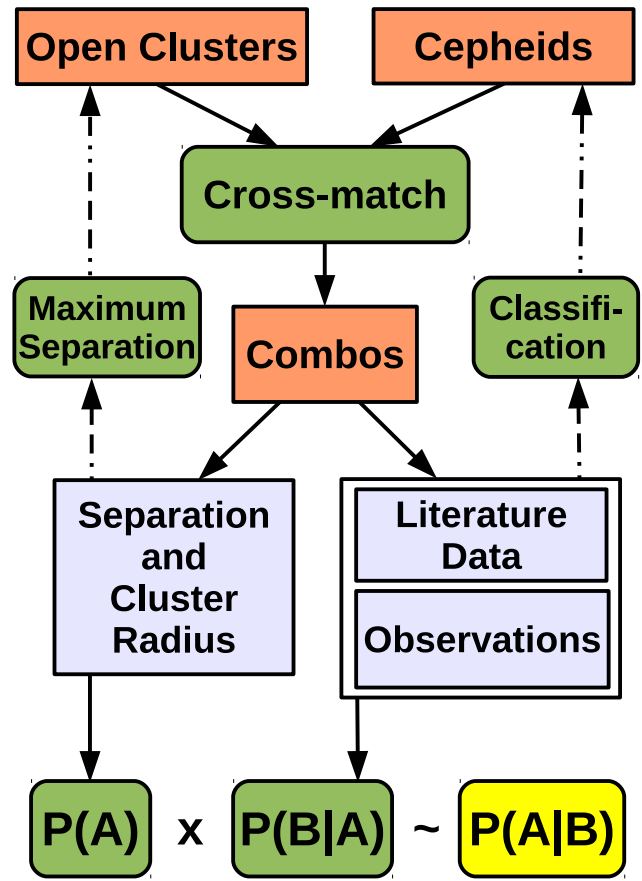}
\caption{Schematic view of membership analysis. Rectangular boxes represent data
sets used, green rounded boxes indicate actions. Cepheids are cross-matched 
(within some maximum separation) with
open clusters to form Combos. Data from the literature and new observations
are combined for each Combo. Cepheid classification is verified based on the
data compiled (light curves, spectra). Priors, $P(A)$, and likelihoods,
$P(B|A)$, are calculated separately and joined as membership probabilities,
$P(A|B)$.}
\label{fig:scheme}
\end{figure}
\label{sec:memcalc}
Our all-sky census is structured as shown in Fig.\,\ref{fig:scheme}. First, 
lists of known open clusters and known
Cepheid candidates are compiled, see Sec.\,\ref{sec:LitData} for details. Second, the two
lists are cross-matched positionally in a many-to-many relationship so that we investigate
a given Cepheid's membership in multiple different open
clusters, and a given open cluster can potentially host multiple
Cepheids. The correct classification of cross-matched Cepheid candidates is verified by considering light curves, and spectra. Misclassified objects are removed from the Cepheids sample.
Third, membership
probabilities are calculated based on all available membership constraints. These
last two points are described in the present section.

Membership probabilities are calculated following Bayes' theorem that 
can be formulated as \citep[\S\,4]{Jaynes2003}:
\begin{equation}
P(A|B)  =  \frac{P(B|A) \times P(A)}{P(B)} \propto P(B|A) \times P(A).
\label{eq:bayes}
\end{equation}
The posterior probability $P(A|B)$ (membership probability)
is proportional to the product of likelihood, $P(B|A)$, and prior, $P(A)$.
$P(B|A)$ represents the conditional probability 
of observing the data under the hypothesis of membership, and 
$P(A)$ quantifies the degree of initial belief in membership. 
The normalization term $P(B)$, of which we possess no knowledge, is the probability to observe the data. We define $P(A)$ in Eqs.\,\ref{eq:prior1} \& \ref{eq:prior2} and $P(B|A)$ in Eq.\,\ref{eq:pba} below.

\subsection{Prior Estimation and
Positional Cross-match}\label{sec:radii} 
\subsubsection{Positional Cross-match}\label{sec:poscrossmatch}
On-sky proximity is a necessary, but insufficient criterion for membership.
Intuitively, if no other information is available, one might tentatively assume
membership for a Cepheid that falls within the core radius of a potential host
cluster.

Therefore, our census starts with a positional cross-match that aims to identify
all combinations of cluster-Cepheid pairs that lie sufficiently close on the sky
to warrant a membership probability calculation (Combos). The cross-match itself
is straightforward: if the separation between a cluster's center coordinates and
the Cepheid's coordinates is smaller than $2.5\,$degrees (to avoid
unnecessary contamination), and less or equal to 5 limiting cluster radii\footnote{This
cut-off radius was adopted to include possible members of cluster halos in the analysis, inspired by the well-known
case of SZ\,Tau in NGC\,1647.}, we  include the Combo in our analysis. 
Using this proximity criterion, we cross-match $\mathbf{990}$ different open
clusters (of $\mathbf{2168}$ in \citealt{Dias}) with $\mathbf{1021}$
Cepheids (of $\mathbf{1821}$ initially compiled) and obtain $\mathbf{3974}$
Combos that we investigate for membership.

The initial cross-match is purely positional, and the majority of Combos studied
are non-members. Our analysis intends to weed out this majority and to indicate
to us the good candidates through a high membership probability.

\subsubsection{The Prior}  
We define the prior, $P(A)$, using the on-sky separation\footnote{We avoid the
term `distance' when referring to the on-sky separation (in arcmin) in order 
to prevent confusion with radial distance (in pc)} between cluster center and
Cepheid, weighted by the cluster radius, i.e. its apparent size on the sky.

The radius of an open cluster is typically determined by fitting an exponential
radial density profile to a stellar over-density on the sky, an approach
originally developed for globular clusters by \cite{1962AJ.....67..471K}. The
method relies on the assumption that two separate distributions are seen: a
constant field  distribution and one that is attributed to the cluster. 

Various ways to define cluster radii can be found in the literature. Among
these are the `core radius' (most stars belong to cluster), $r_c$, and
the `limiting radius' for the cluster halo (strong field star contamination),
$r_{\rm lim}$, see
\citet{2005A&A...438.1163K,2005A&A...440..403K} and \citet{Buko11}. 

Intuitively, the probability of membership is related to separation and cluster radius, cf. \cite{2010A&A...510A..78S}. Let us therefore define the quantity $x$ as:
\begin{equation}
x = \frac{r - r_c}{2r_{\rm lim} - r_c}  ,
\label{eq:prior0}
\end{equation}
where $r$ denotes separation. $x$ is negative,
if the Cepheid lies within the cluster's (projected) core and becomes unity at a
separation equal to twice the limiting radius.
We define our prior, $P(A)$, so that (no other constraints considered)
membership is assumed when the Cepheid lies within the cluster's core, i.e. $x < 0$.
Outside $r_c$, inspired by radial density profiles of star clusters, we let the prior
fall off exponentially and define it to reach $0.1\% = 10^{-3}$ at $x =
1$.
Hence,
\begin{equation}
P(A)(x < 0) \equiv 1
\label{eq:prior1}
\end{equation}
\begin{equation}
P(A)(x \geq 0) \equiv 10^{-3x} \rm{\hspace{0.1cm}}.
\label{eq:prior2} 
\end{equation}
\begin{figure}
\centering
\includegraphics{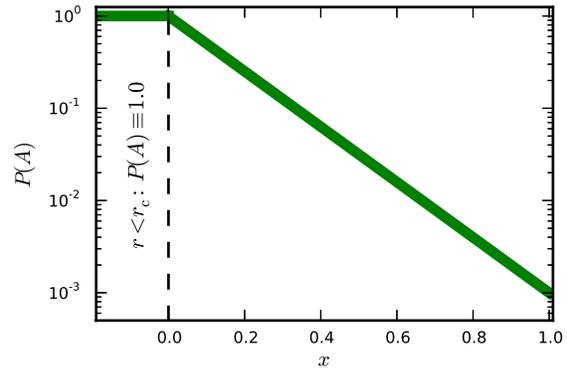}
\caption{Illustration of the adopted prior, $P(A)$, as a function of separation normalized to a cluster's radii, expressed by the quantity $x$, cf. Eq.\,\ref{eq:prior0}. If $x \leq 0$ (Cepheid within core radius, $r_{\rm{c}}$): $P(A) \equiv 1$. Outside the core, $P(A)$ decreases exponentially, inspired by radial density profiles of stars clusters. We adopt $P(A) = 0.001$ at $r = 2\,r_{\rm{lim}}$.}
\label{fig:prior}
\end{figure} 
Figure\,\ref{fig:prior} serves to illustrate this definition. The prior thus carries the 2-dimensional information of separation and cluster radius, and thereby takes into account how concentrated a cluster is on the sky assuming circularly distributed member stars.

\subsection{The Likelihood $\mathbf{P(B|A)}$}\label{sec:likelihood}
The likelihood, $P(B|A)$, is computed as a hypothesis test. It 
estimates the probability that the observed data is consistent with the null hypothesis of (true) membership. This approach was inspired by the
Hipparcos astrometry-based studies by \cite{1999A&A...345..471R} and \cite{2000A&AS..146..251B}. We extend it here to take into account up
to $6$ dimensions using parallax, $\varpi$, radial velocity (RV), proper motion, $\mu_\alpha^*$ and $\mu_\delta$, iron abundance, [Fe/H], and age (open clusters assumed to be co-eval), weighting all constraints equally.  

Assuming that a given Cepheid was not used to determine a cluster's 
(mean) parameters, we can calculate the quantity 
\begin{equation}
c = x^T \Sigma^{-1} x,
\end{equation}
where $x$ denotes the vector containing as elements the differences between the (mean) cluster and Cepheid quantities:
\begin{equation}
x = \left( \varpi_{\rm{Cl}} - \varpi_{\rm{Cep}}, \langle v_{r,\rm{Cl}} \rangle - v_{\gamma,\rm{Cep}}, \ldots \right).
\label{eq:xvec}
\end{equation}
Let $\rm \mathbf{C}_{Cl}$ be the covariance matrix of the cluster and $\rm \mathbf{C}_{Cep}$ that of the Cepheid. Let $\mathbf{\Sigma}$ then denote the sum of the two and $\mathbf{\Sigma}^{-1}$ its inverse.
Since the data employed in this calculation comes from many different sources, no knowledge of correlations between the different parameters is available. We thus make the assumption of independent measurements, which results in diagonal covariance matrices containing only parameter variances. Possible correlations between Cepheid and Cluster parameters are thus assumed to be negligible. We consider this justified, since we possess no knowledge of the extent of such correlations and assume that Cepheids were not used in the determination of cluster mean values. This formulation furthermore implicitly assumes normally (Gaussian) distributed errors.

Under these assumptions $c$ is $\chi^{2}$ distributed, i.e. $c \sim \chi^{2}_{N_{\rm dof}}$, where 
$N_{\rm dof}$ is the number of degrees of freedom equal to the length of
vector $x$, ranging from $1$ to $6$. $c$ thus depends on the number of membership constraints considered  (the on-sky position is used in the prior). In cases where no membership constraints are available, i.e. $N_{\rm dof} = 0$, we set $P(B|A) \equiv 1.$

$P(B|A)$ is obtained by calculating unity minus the p-value of c, ${\rm p}(c)$:
\begin{equation}
	P(B|A) = 1 - {\rm p}(c).
	\label{eq:pba}
\end{equation}
Since the $\chi^2$ distribution (and therefore the p-value computed) is very sensitive to $N_{\rm dof}$ for small $N_{\rm dof}$, $P(B|A)$ naturally contains information on the number of membership constraints employed.

Of course, we cannot prove the null hypothesis, only exclude it. However, by including the greatest number of the most stringent membership constraints possible,
this method very effectively filters out non-members. The remaining
candidates can therefore be considered \textit{bona-fide} members, provided the
constraints taken into account are sufficiently strong.

The filtering effectiveness of the likelihood strongly depends on the
uncertainties adopted for the constraining quantities: the larger the error, the
weaker the constraint. Conversely, the smaller the error, the more important become systematic differences between quantities measured or inferred through different techniques. Obtaining reasonable estimates of the external uncertainties is of paramount importance to the success of this work, since the data considered is inhomogeneous and listed uncertainties typically provide formal errors or estimates of precision. 

For certain quantities, we therefore adopt increased error budgets that we motivate and detail in the following sections. Care is taken to avoid too large or too small error budgets, and to ensure that likelihood remains an effective membership criterion.

\section{Data Used To Compute Likelihoods}\label{sec:Data}
In this section, we describe how we compile the data used for our analysis.
The constraints employed are: on-sky separation, parallax, proper motion, radial velocity, and the population parameters iron abundance
(as a proxy for metallicity) and age. Most data considered originates from published literature and catalogs. However, we also include radial velocity (RV) data from an extensive, year-round observation program carried out on both hemispheres. Some details on this program are provided in Sec.\,\ref{sec:CepheidRV}. A full description, however, is out of scope for this work and will be published separately.

Very often, data on a given membership constraint can be found in different
references. In such cases, a choice of which reference to prefer over the other
ones has to be made. In each of the following subsections, the references
mentioned first are the ones preferentially adopted. This section is divided
into two parts: Sec.\,\ref{sec:OCcompilation} dedicated to open clusters, and
Sec.\,\ref{sec:Cepcompilation} to Cepheids. Stellar associations are not
considered.

\subsection{Open Cluster Data}\label{sec:OCcompilation}
For open cluster data used in this work, we largely
rely on the \citet[from hereon: D$02$]{Dias} catalog\footnote{Version V3.3,
16 January 2013}, which builds partially on the WEBDA
database\footnote{Maintained by E. Paunzen and C.
St\"utz in Vienna, cf. \texttt{http://http://www.univie.ac.at/webda/}}
originally developed by \cite{1988BICDS..35...77M,1995ASSL..203..127M}, where
additional useful information, e.g. on radial velocities, can be found.
D$02$ is an extensive ``living'' compilation of open cluster data that is
regularly updated with the latest available information on open clusters. Thanks to
this process, we can assume that the most accurate available information
available for the open clusters is used in our analysis. D$02$ is
furthermore the most complete compilation of open clusters available, so that
only few potential Cepheid host clusters are missed (cf. Sec.\,\ref{sec:missedCCs}).

The definition of $P(A)$ in Eqs.\,\ref{eq:prior0} through
\ref{eq:prior2} requires information for two types of radii, core and limiting. Since D$02$
lists only a single quantity, apparent diameters, we adopt core and limiting
radii from other sources, see Sec.\,\ref{sec:Clradii}. Further choices made
regarding cluster data are presented in the subsections concerning
parallax (Sec.\,\ref{sec:Clplx}), proper motion (Sec.\,\ref{sec:Clpm}), mean
radial velocity (Sec.\,\ref{sec:ClRV}), iron abundance (Sec.\,\ref{sec:ClFeH}),
and age (Sec.\,\ref{sec:ClAge}).

\begin{figure*}
\includegraphics{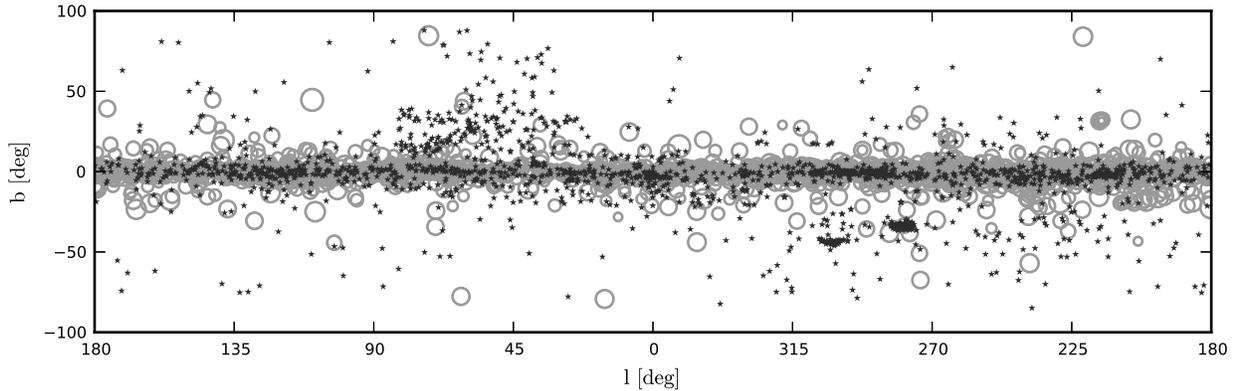}
\caption{Distribution of open clusters and Cepheids compiled, shown in Galactic
coordinates. Light gray open circles represent open clusters with
markers logarithmically scaled for apparent size, darker gray star symbols 
Cepheids.}   
\label{fig:Clcoord}
\end{figure*}

Figure\,\ref{fig:Clcoord} shows the distribution of clusters (black open
circles, scale with limiting radius) and Cepheids (light star symbols) in 
Galactic coordinates. Clusters closely trace the disk, and no obvious gaps are
present in our all-sky census.
\label{sec:LitData}

\subsubsection{Cluster radii}\label{sec:Clradii}
In order to choose which literature radii to adopt, we start by
investigating to what degree cluster radii are reliable quantities. To this end, we search the
literature for extensive catalogs that provide both core and limiting radii. 
Three such studies are identified:
\citet[from hereon, we refer to the combined catalog from both studies as K05]{2005A&A...438.1163K,2005A&A...440..403K},
\citet[from hereon: B11]{Buko11}, and \citet[from hereon:
K12]{2012A&A...543A.156K}. We do not include \cite{2007MNRAS.374..399F} here, since we notice a suspicious correlation
between $r_c$ and $r_{\rm lim}$. Note, however, that some clusters listed in
K$12$ were originally identified by \cite{2007MNRAS.374..399F}. 

\begin{figure}
\centering
\includegraphics{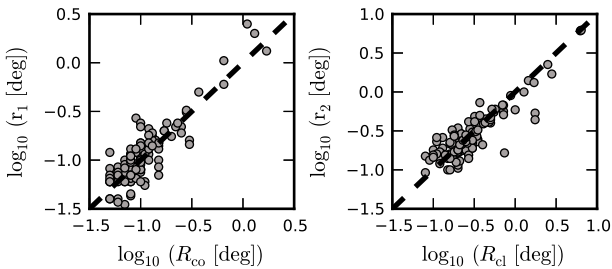}
\caption{\textit{Left panel:} comparison between $r_1$ in K$12$ and the core
radius, $R_{\rm{co}}$, in K$05$; \textit{right panel:} same for $r_2$ in K$12$
and the limiting radius, $R_{\rm{cl}}$, in K$05$. The radii are comparable.}
\label{fig:radcompare}
\end{figure} 
Since many clusters in K$12$ were also studied by K$05$, we compare the
three radii defined in K$12$ with the core and limiting radii in K$05$ and notice
that the limiting radius in K$05$, $R_{\rm{cl}}$, corresponds well to $r_2$ in
K$12$ (though $r_2$ tends to be smaller), while $r_1$ in K$12$ is rather similar to the core
radius in K$05$, $R_{\rm{co}}$. Nevertheless, a fair amount of scatter exists
between both studies, see Fig.\,\ref{fig:radcompare}. We consider K$12$ an update (and extension) 
of K$05$ and therefore prefer the newer cluster parameters over the older ones.

We previously compared radii given in K$05$ and B$11$ for the clusters
common to both works in \cite{2012IAUS..285..275A}. Rather large scatter is present (more than a
factor of $2$ for an appreciable fraction) and illustrates that cluster radii are
subject to significant uncertainty. However, the radii from both studies
follow the same trend and we therefore consider them comparable for our purpose,  
although K$05$ and K$12$ are based on optical and B$11$ on near-infrared (NIR)
$2$MASS \citep{2003tmc..book.....C} photometry.

Given the sometimes rather large difference between cluster radii
mentioned in the literature, we adopt a `permissive' scheme that gives
preference to the study giving the largest limiting radius for the cluster 
and thereby bias ourselves towards higher $P(A)$. We therefore
strongly rely on the remaining membership constraints that define the likelihood
to filter out chance alignments.

\begin{figure}
\centering
\includegraphics{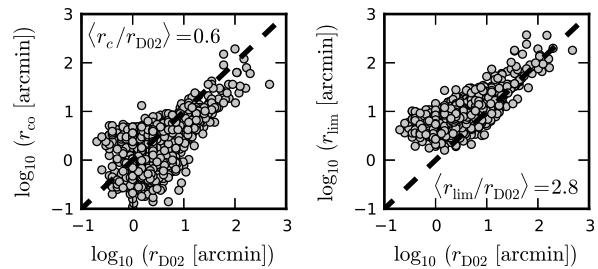} %
\caption{Radii based on apparent diameters from D02 compared to the core
(left panel) and limiting radii (right panel) compiled from K12, K05, and B11,
see text. Median ratios printed on the graph.}
\label{fig:radcomparedias}
\end{figure}
For $478$ clusters cross-matched, only an apparent diameter was
available.
For these, we thus approximate $r_{\rm c}$ and $r_{\rm lim}$ from the typical
(here: median) ratios of D02 apparent radii, $r_{\rm{D02}}$, and the $r_c$ and
$r_{\rm lim}$ adopted as described above. Figure\,\ref{fig:radcomparedias}
illustrates this: the median ratio of $\langle r_{\rm{c}} / r_{\rm{D02}}
\rangle = 0.6$, and $\langle r_{\rm lim} / r_{\rm{D02}} \rangle = 2.8$. 
Priors estimated using this approach are identified in the online tables
and marked with an asterisk in the tables presenting our results.

\subsubsection{Cluster parallax}\label{sec:Clplx}
Distances in [pc], listed in D02, are converted to parallaxes in [mas]
through Eqs. \ref{eq:cepplx} and \ref{eq:cepplxerror}, see Sec.
\ref{sec:CepheidDistances}. Since most cluster distances listed in D02 are based
on isochrone-fitting, i.e., are model dependent, we adopt an error
budget of $20\%$ to account for uncertainties arising from rotation, binarity,
metallicity, and other modeling-related effects.

\subsubsection{Mean Proper Motion}\label{sec:Clpm}
Mean cluster proper motions, $\bar\mu_{\alpha,Cl}^*$ and $\bar\mu_{\delta,Cl}$,
are provided in D$02$.

The uncertainties on mean proper motion listed in these references are typically
calculated either as intrinsic dispersions (e.g. for clusters closer than
approx. 400pc originally studied in K$12$), or as standard mean errors,
i.e.
the error decreases as $\sqrt{N_{*}-1}$, where $N_{*}$ is the number of stars considered members, cf.
D$02$\footnote{See under `version 2.3 (25/abr/2005)' in file:
\texttt{http://www.astro.iag.usp.br/$\sim$wilton/whatsnew.txt}}. The quoted
uncertainties on the cluster mean are thus much smaller than the uncertainty on
an individual cluster star's measurement. For example, in K$12$, the typical
mean proper motion error is 0.4 mas\,yr$^{-1}$.

For the majority of Cepheids, however, the uncertainties on proper motion are
much larger, and many have been obtained from different data sets, using
different techniques. Therefore, to ensure comparability of inhomogeneous data
and to reduce our sensitivity to offsets in zero-points due to data-related
specificities such as reduction techniques, we adopt a more generous error
budget for $\bar\mu_{\alpha,Cl}^*$ and $\bar\mu_{\delta,Cl}$ that resembles the
uncertainty of an individual cluster star's proper motion. This is done by
multiplying the uncertainty listed by the factor $\sqrt{N_{*}-1}$ and thus
slightly reduces the weight of proper motion as a membership constraint.
Empirically, we are confident that this is justified, since proper motions of
Cepheids typically barely exceed their uncertainties, and care should be taken
not to over-interpret their accuracy.

\subsubsection{Mean Radial Velocity}\label{sec:ClRV}
\begin{figure}
\includegraphics{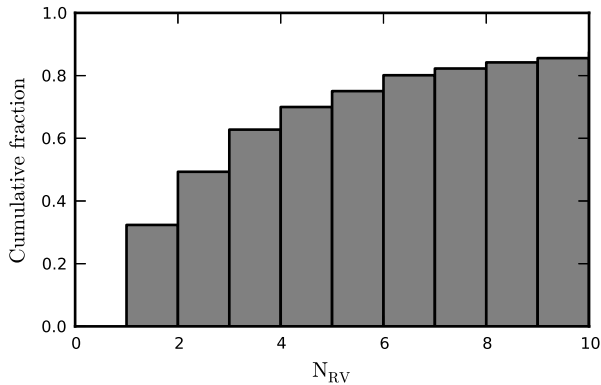}
\caption{Cumulative fraction of clusters for which a given number of stars,
$N_{\rm RV}$, was used to determine the average RV. About half of average cluster RVs are based on a couple of stars.}
\label{fig:disc_rv}
\end{figure}

Average cluster radial velocities and associated errors are
listed in D02.
However, qualitative differences can exist in the uncertainties listed.
For some well-studied clusters, the uncertainty given is an estimate of the intrinsic RV
dispersion. For the majority of cluster RVs, however, only a few stars were used
to determine the mean cluster RV (about half on two stars or less, cf.
Fig.\,\ref{fig:disc_rv}). These cases are therefore subject to systematic
uncertainties due to implicit membership assumptions, for instance. In
addition, unseen binary companions and instrumental zero-point offsets can introduce
systematic uncertainties at the level of a few km\,s$^{-1}$.

We therefore adopt $2$\,km\,s$^{-1}$ as a minimum uncertainty of the mean cluster velocity. If no uncertainty estimate is given, we adopt $\sigma(RV_{\rm Cl}) = 10$\,km\,s$^{-1}$ / $\sqrt{N_{RV}}$ as a typical uncertainty on the mean cluster RV, where $N_{RV}$ is the number of stars used to determine the mean cluster RV.

\subsubsection{Iron Abundance}\label{sec:ClFeH}
We adopt iron abundances compiled in D$02$, including the uncertainties given.
The mean uncertainty among the clusters compiled is $0.08$\,dex. 

\subsubsection{Cluster Age}\label{sec:ClAge}
Ages were available for most clusters, since they are often
determined simultaneously with the distance via isochrone-fitting. 
Although a model-dependent parameter, age does provide a valid constraint for
membership, reflecting evolutionary considerations that are
empirically validated. Quantifying an uncertainty for age as a parameter,
however, is rather difficult.

Younger clusters exhibit a Main Sequence turn-off at higher stellar
masses than older clusters. As a consequence of the Initial Mass
Function, a younger cluster's turn-off point tends to be less
populated than that of an older cluster. It therefore follows that age
estimates tend to become more accurate with age, since the cluster's turn-off
point tends to be defined more clearly against the field and therefore better
constrains an isochrone fit.

\begin{figure}
\centering
\includegraphics{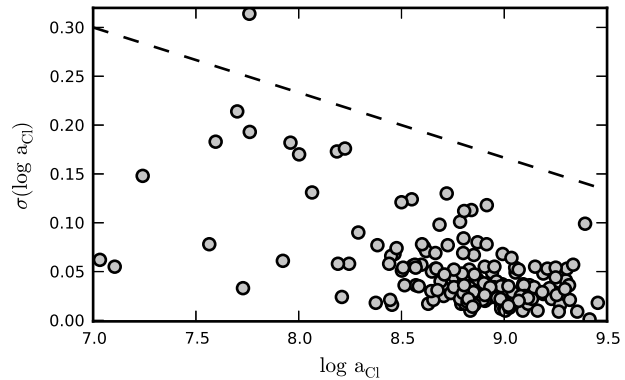}
\caption{Age uncertainty, $\sigma(\log{\rm{a}_{Cl}})$, as function of cluster
age, $\log{\rm{a}_{Cl}}$, given by K$12$. Older clusters have more precisely
estimated ages. We adopt as error budget for clusters without stated age
uncertainties an upper limit to this proportionality indicated by the dashed
line, cf. Eq.\,\ref{eq:ClAgeError}.}
\label{fig:clages}
\end{figure}

Figure\,\ref{fig:clages} corroborates the above reasoning by showing cluster ages against their uncertainties as given in K$12$. We thus estimate an upper limit on the uncertainty of cluster age as the dashed line in Fig.\,\ref{fig:clages}, which is:
\begin{equation}
\sigma(\log{\rm{a}_{Cl}}) \leq 0.3 - 0.067 \left( \log{\rm{a}_{Cl}} - 7.0  \right)\,.
\label{eq:ClAgeError}
\end{equation}

\subsection{Cepheid Data}\label{sec:Cepcompilation}
Cepheid candidates were compiled from the
January 2012 version of the General Catalog of Variable Stars \citep[from
hereon: GCVS]{GCVS} and the May 2012 version of the AAVSO Variable Star Index
(from hereon: VSX)\footnote{\texttt{http://www.aavso.org/vsx/}}. From GCVS and
VSX, we import the variability types CEP, CEP(B), DCEP, DCEPS; from VSX, we
include the ASAS \cite{1997AcA....47..467P,2002AcA....52..397P,2005AcA....55..275P} Cepheid candidates classified as DCEP-FU or DCEP-FO. This list also contains Cepheid candidates found by ROTSE \citep{2000AJ....119.1901A} or NSVS \citep{2004AJ....127.2436W}, as well as the ones in the suspected variables catalog \citep{1982ncsv.book.....K}.

This starting point contains an unknown, but probably high, fraction of
non-Cepheids. Type-II Cepheids (halo objects) and Cepheids belonging to
the Magellanic Clouds are mostly removed from the sample by cross-matching with the
clusters (trace the disk, see Fig.\,\ref{fig:Clcoord}). 
To further reduce contamination, we visually inspect all ASAS-$3$ V-band light
curves of Cepheid candidates with ASAS identifiers.

Radial pulsation and color variations during the pulsation are defining
characteristics of Cepheids. We thus use the spectra obtained for radial
velocity observations described in Sec.\,\ref{sec:CepheidRV} to verify
classification. A total of $\mathbf{151}$ ASAS Cepheid candidates and
$\mathbf{32}$ others are thus rejected from the Cepheid sample, resulting in a
final list of 1821 Cepheid candidates, $\mathbf{1021}$ of which are
cross-matched with open clusters.

The cleaned sample of Cepheids cross-matched with clusters was appended with
literature data from many sources, and references are given in the text. Among the most relevant references are:
\begin{itemize}
  \item The \cite{1995IBVS.4148....1F} DDO Cepheid
  database\footnote{\texttt{http://www.astro.utoronto.ca/DDO/research/Cepheids/}}
  \item The \cite{2009A&A...504..959K} Cepheid database (KS$09$)
  \item The ASAS Catalog of Variable Stars \citep[ACVS]{2005AcA....55..275P} and associated photometry
  \item The new Hipparcos reduction \citep{2007ASSL..350.....V}
  \item The extended Hipparcos compilation \citep[XHIP]{2012AstL...38..331A}
  \item The ASCC-2.5 catalog \citep{2001KFNT...17..409K} updated by
  \cite{2007AN....328..889K}
  \item The PPMXL catalog \citep{2010AJ....139.2440R} 
  \item The $2$MASS catalog \citep{2003tmc..book.....C}
  \item The Cepheid photometry obtained by
  \cite{2000A&AS..143..211B,2008yCat.2285....0B}
  \item The McMaster Cepheid photometry and radial velocity data archive
  maintained by Doug Welch\footnote{\texttt{http://crocus.physics.mcmaster.ca/Cepheid/}}
  \item The radial velocity data, see Sec.\,\ref{sec:CepRV}.
\end{itemize}

\subsubsection{Cepheid Parallaxes}\label{sec:CepheidDistances}
Parallax, $\varpi$, is a key membership constraint, since cluster membership is
virtually guaranteed if a Cepheid occupies the same space volume as a cluster.
We combine parallax estimations from different sources, favoring
\textit{PLR-independent} determinations.

Parallax in [mas] is given preference over distance in [pc] here, since the
uncertainty, $\sigma_\varpi$, is normally distributed, in contrast to the error
in distance. This is important, since the computation of likelihoods by Eq.
\ref{eq:pba} assumes Gaussian uncertainties.

We compile parallaxes from \citet[8 Cepheids]{2007AJ....133.1810B},
\citet[65 Cepheids]{2011A&A...534A..94S}, and the new Hipparcos reduction by \citet[so long as $\sigma_\varpi / \varpi \leq 0.1$ and
$\varpi > 0$, 5 Cepheids]{2007ASSL..350.....V}. We then calculate PLR-based
parallaxes for 622 additional Cepheids, see below.

PLR-based parallaxes of fundamental-mode Cepheids are calculated from distances
computed following \cite{2010OAP....23..119T}. Our choice of P-L relation was
motivated mainly by the considerations that i) V-band magnitudes can be obtained
for the largest number of Cepheids; ii) the above formulation is calibrated for
the Galaxy using the most recent observational results, including the HST
parallaxes by \cite{2007AJ....133.1810B} and the cluster Cepheids from
\cite{2010Ap&SS.326..219T}.

We thus calculate PLR distances as follows:
\begin{equation}
5 \log{d} = \langle m_V \rangle - \langle M_V \rangle - A_V + 5 \,, 
\label{eq:distCep}
\end{equation}
where $\langle m_V \rangle$ is the apparent mean V-band magnitude, and the average absolute V-band magnitude, $\langle M_V \rangle$, is obtained from the pulsation period $P$ via:
\begin{equation}
\langle M_V \rangle = - \left( 1.304 \pm 0.065 \right) - \left( 2.786 \pm 0.075 \right) \log{P} \,.
\label{eq:plr}
\end{equation}
Eq.\,\ref{eq:plr} is valid only for fundamental-mode pulsators, no distances were estimated for overtone pulsators. 
The total absorption, $A_V$, is defined as:
\begin{equation}
A_V = R_V \cdot E(B-V)\,,
\label{eq:anu}
\end{equation}
with $R_V = 3.1$ the canonical ratio of total to selective extinction (reddening
law) and $E(B-V)$ the color excess of the object, cf. Sec.\,\ref{sec:ebv}.
The parallax is simply:
\begin{equation}
\varpi = \frac{1000}{d}\,\rm{[mas]}\,
\label{eq:cepplx}
\end{equation}
with $d$ in [pc]. The parallax uncertainty, $\sigma_\varpi$, is obtained considering the error budget on the distance, $\sigma_d$:
\begin{equation}
\sigma_\varpi = \frac{1000}{d^2}\cdot \sigma_d\,\rm{[mas]}\,.
\label{eq:cepplxerror}
\end{equation}

Thus, to estimate a Cepheid's parallax, knowledge of the PLR, $P$, $\langle m_V
\rangle$, and $A_V$ is required. Periods are usually available in the GCVS or the VSX, whereas average magnitudes and color excesses of many of the newer Cepheid candidates are not available in the literature. However, $E(B-V)$ can be estimated from combined NIR and optical data. The following paragraphs describe in detail how these quantities are compiled.

\begin{figure}
\includegraphics{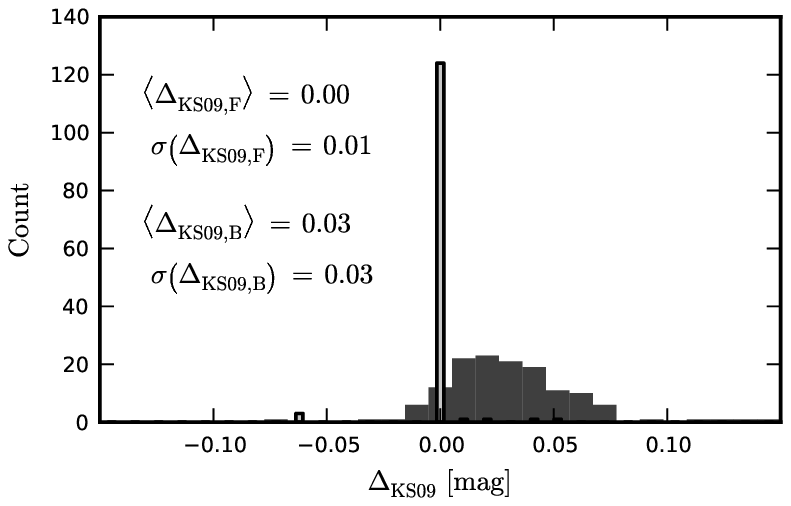}
\caption{Histogram of differences in mean magnitudes relative to the KS$09$
values, $\Delta_{\rm{KS09}}$. Light slim bars show
$\Delta_{\rm{KS09,F}}$ computed using Fernie values (123 Cepheids); Darker
broad bars with no outline show $\Delta_{\rm{KS09,B}}$ computed using data
from \citealt{2000A&AS..143..211B} (127 Cepheids). Mean differences, $\langle \Delta_{\rm{KS09}} \rangle$, and
dispersions, $\sigma\left(\Delta_{\rm{KS09}}\right)$, given in [mag].}      
\label{fig:deltamag_KS9vsBERDhist}
\includegraphics{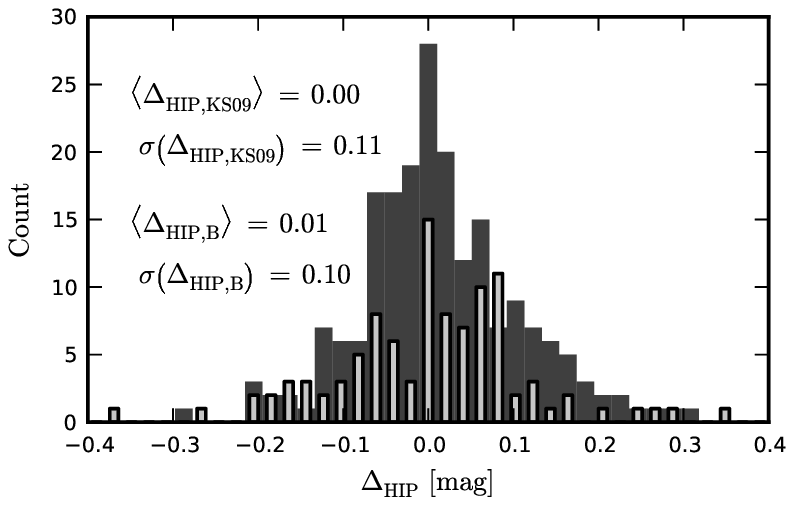}
\caption{Histogram of differences in mean magnitudes relative to Hipparcos
median V-band, $\Delta_{\rm{HIP}}$. Light slim bars show
$\Delta_{\rm{HIP,KS09}}$ computed using KS$09$ (104 Cepheids). Darker broad
bars with no outline show $\Delta_{\rm{HIP,B}}$ computed using
\citealt{2000A&AS..143..211B} (198 Cepheids). Mean differences, $\langle \Delta_{\rm{HIP}} \rangle$, and dispersions,
$\sigma\left(\Delta_{\rm{HIP}}\right)$, given in [mag].}     
\label{fig:deltamag_XHIPvsBERDhist}
\includegraphics{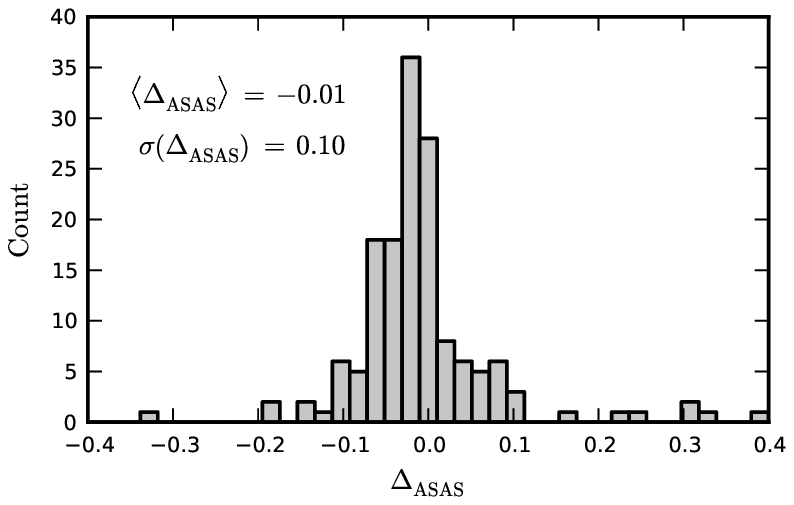}
\caption{Histogram of differences in mean V-band magnitude computed from ASAS
light curves for 154 Cepheids relative to reference values from Fernie and
KS$09$, $\Delta_{\rm{ASAS}}$, cf. also Fig.\,\ref{fig:deltamag_KS9vsBERDhist}.
Mean difference, $\langle \Delta_{\rm{ASAS}} \rangle$, and dispersion,
$\sigma\left(\Delta_{\rm{ASAS}}\right)$, given in [mag].}
\label{fig:deltamaghist}
\end{figure}
\paragraph{Mean Magnitude, $\mathbf{\langle m_V \rangle}$}\label{sec:optimizeP} 
We compile mean V-band magnitudes, $\langle m_V \rangle$, from multiple
references. Different methods of determining mean magnitudes exist, and the
photometry employed is inhomogeneous, forcing us to adopt a zero-point for mean
magnitudes compiled. In Fig.\,\ref{fig:deltamag_KS9vsBERDhist} we therefore
compare mean magnitudes from \citet[from hereon: KS$09$]{2009A&A...504..959K}
with the  \cite{1995IBVS.4148....1F} database's magnitude-based means and the
intensity-means from \cite{2000A&AS..143..211B}. For the Cepheids common to both
studies, KS$09$ and the Fernie magnitudes show excellent agreement. We therefore
adopt the following order of preference for compiling mean magnitudes.

First, we adopt $\langle m_V \rangle$ values from KS$09$ with a fixed error
budget of $0.03$\,mag, since the study carefully investigates amplitudes with a
special focus on binarity.

Second, we adopt the magnitude-based means from the Fernie database with
uncertainties calculated as the difference between intensity- and magnitude-mean
magnitudes, with a minimum error of $0.03$\,mag.

Third, we include \cite{2000A&AS..143..211B} mean magnitudes. As seen in
Fig.\,\ref{fig:deltamag_KS9vsBERDhist}, these $\langle m_V \rangle$ values are
systematically smaller (brighter) by approx. $0.03$\,mag than KS$09$. This
discrepancy is most likely due to different ways of determining the mean. We
remove this offset from the \cite{2000A&AS..143..211B} values for internal
consistency and adopt $0.03$\,mag as error budget for these values, identical to
KS$09$.

Fourth, we employ median V-band magnitudes from the Hipparcos catalog
(\citealt{1997ESASP1200.....P}, obtained via XHIP) for 8 Cepheids.
The median V-band magnitudes derived from Hipparcos magnitudes can differ
significantly from mean magnitudes listed in other references, cf.
Fig.\,\ref{fig:deltamag_XHIPvsBERDhist}. Usually, this is due to contamination
due to a nearby star within the instantaneous field of view.
As error budget for the Hipparcos median V-magnitudes, we adopt
$sigma(\Delta_{\rm{HIP,KS09}}) = 0.110$\,mag, see 
Fig.\,\ref{fig:deltamag_XHIPvsBERDhist}. We note that we
could find no dependence on period or number of transits for this dispersion.

Fifth, we adopt average apparent V-band magnitudes that we determine from ASAS-3
light curves.
To this end, we fit Fourier series (same procedure as described for radial
velocities in Sec.\,\ref{sec:vgamma}) to the phased light curves and use the
constant term as the average, $\langle m_V \rangle$. Figure
\ref{fig:deltamaghist} shows a histogram of $\Delta_{\rm ASAS}$, the differences
between the computed ASAS-based $\langle m_V \rangle$ and Fernie or KS$09$.
We remove the offset of $-0.01$\,mag from the ASAS mean magnitudes and adopt the
dispersion of $0.10$\,mag computed as the error budget.

\begin{figure*}
  \centering
  \includegraphics{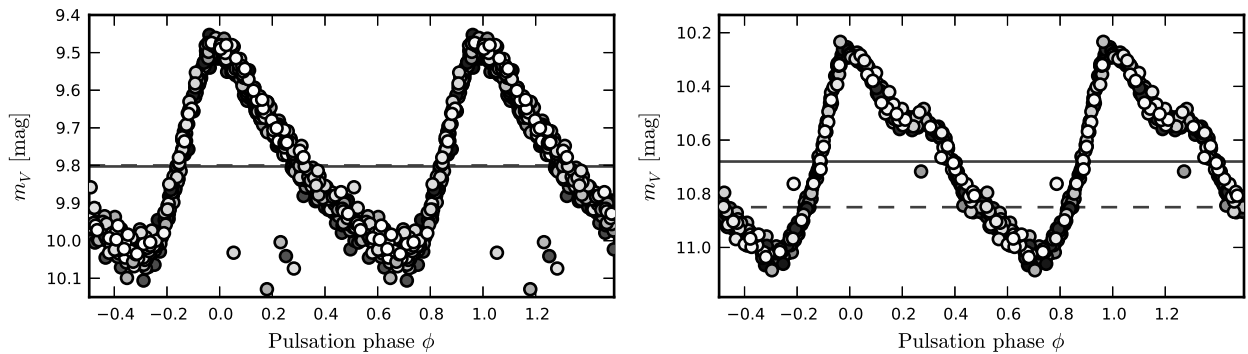}
  \caption{Phase-folded ASAS-3 V-band light curves of CY\,Car (left) and BM\,Pup
  (right). Julian date of observation indicated in grayscale, 
  increasing from black to white. Horizontal lines indicate reference average
  magnitude (dashed, KS$09$) and constant term of the fitted
  Fourier series (solid). For CY\,Car, the two are in excellent agreement. 
  BM\,Pup has a bright neighbor that contaminates the aperture used to measure its flux, 
  leading to an underestimated (too bright) mean magnitude.}
  \label{fig:meanV}
\end{figure*}

The large dispersion, $\sigma\left(\Delta_{\rm ASAS}\right)$, in Fig.
\ref{fig:deltamaghist} probably originates from contamination due to nearby
stars.
To illustrate this, Fig.\,\ref{fig:meanV} shows phase-folded ASAS-3 V-band
light curves of two Cepheids, CY\,Car (left) and BM\,Pup (right).
Our mean magnitude agrees well with the literature value for CY\,Car. For
BM\,Pup however, a systematic difference of approximately $0.144$\,mag is
evident, although the light curve appears to be clean otherwise. Inspection of a
DSS images, however, reveals that contamination from a nearby companion is
likely.
Out of 154\,Cepheids for which the ASAS light V-band curves were inspected, 20
differed by more than 0.1\,mag from the reference value, and 28 agreed to within
0.01\,mag.

If no mean magnitude is obtained from any of the above sources, we perform a
(rough) estimate of $\langle m_V \rangle$ based on the information provided in
the GCVS and the VSX, using the magnitude at maximum brightness, $\mbox{min}_V$,
and the amplitude, $\mbox{amp}_V$, of the V-band light curve. $\mbox{amp}_V$ is
either provided directly by the catalogs, or calculated as the difference
between minimum and maximum brightness, $\mbox{amp}_V = \mbox{max}_V -
\mbox{min}_V$.

\begin{figure}
\centering
\includegraphics{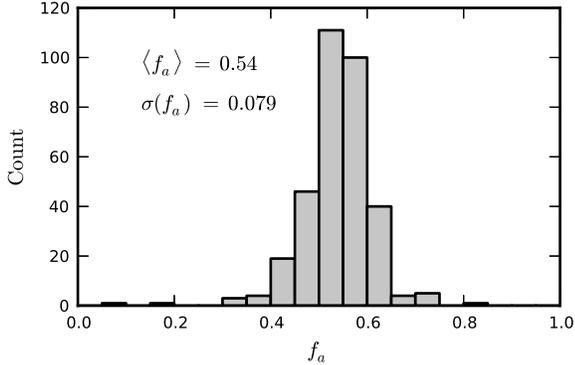}
\caption{Fractional amplitudes, $f_a$, computed from Fernie mean, and GCVS maximum and minimum magnitudes, see Eq.\,\ref{eq:ratioamp}.}
\label{fig:fractionalamp}
\end{figure}

Since Cepheid light curves are skewed, their mean magnitudes do not necessarily
lie at half the amplitude. We therefore estimate the typical fractional
amplitude at mean brightness, $\langle f_a \rangle$, to compute $\langle m_V
\rangle = \mbox{min}_V + \langle f_a \rangle\,\mbox{amp}_V$.
Figure \ref{fig:fractionalamp} shows a histogram of $f_a$ computed using mean
magnitudes listed in the Fernie database, $\langle m_{V,{\rm F}} \rangle$, and
amplitudes, $\mbox{amp}_V$, from the catalogs.
We find
\begin{equation}
\langle f_a  \rangle \equiv \mbox{median} \left( \frac{\langle m_{V,{\rm F}} \rangle -\mbox{max}_V }{\mbox{min}_V - \mbox{max}_V} \right) = 0.54 \pm 0.079\,.
\label{eq:ratioamp}
\end{equation}
We derive an uncertainty on $\langle m_V \rangle$ thus obtained using the uncertainty on $\langle f_a  \rangle$, an estimated error on the amplitude, and a prescribed error on the magnitude at maximum brightness (we adopt $0.1$\,mag for 12th magnitude and below, and increase linearly to $0.5$\,mag at 20th magnitude). The resulting mean error on $\langle m_V \rangle$ is $0.27$\,mag.

This estimation works reasonably well, although there exist obvious limitations,
such as inhomogeneity of pass-bands, accuracy of the upper and lower limits, the
applicability of the above ratio for a given Cepheid. Nevertheless, it does
provide access to rough estimates of $\langle m_V \rangle$ for Cepheids with
little available information.

Given the many different ways in which average magnitudes were estimated, we
keep track of the type of estimation to ensure traceability of any potential
issues.

\paragraph{Color excess, $\mathbf{E(B-V)}$} \label{sec:ebv}
\begin{figure}
\centering
\includegraphics{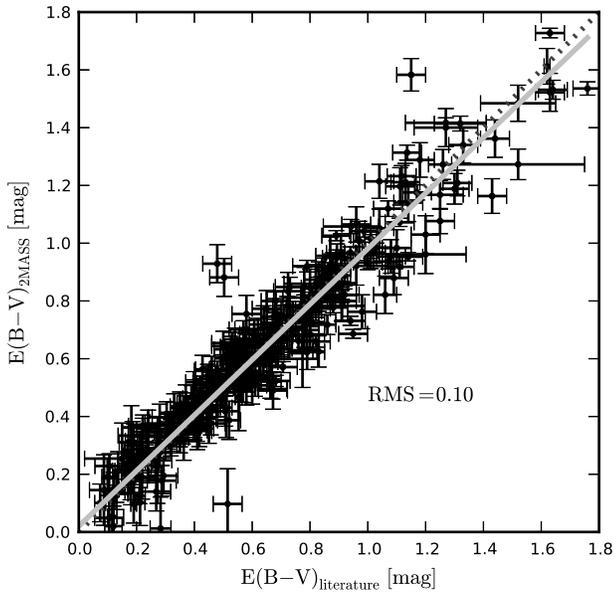}
\caption{Color excess from the literature against 2MASS-based estimate.
The solid line represents a weighted
least squares fit, a dashed line indicates the diagonal. 
RMS around either line: 0.10 mag.}
\label{fig:cepebv}
\end{figure}
The principal references adopted for color excess are
\cite{2008MNRAS.389.1336K,2007MNRAS.377..147L,2007A&A...473..579S,2007A&A...476...73F}.
Where available, we adopt stated uncertainties, $\sigma\left(E(B-V)\right)$.
If no $\sigma\left(E(B-V)\right)$ are listed, we adopt an error budget
of $0.05$\,mag for
\cite{2008MNRAS.389.1336K} and \cite{2007A&A...473..579S}, 
and $0.03$\,mag for \cite{2007MNRAS.377..147L}. 

For other Cepheids , we adopt $E(B-V)$ from the Fernie database and an
error budget of $0.05$\,mag, unless the standard error from multiple reddening
estimations was given.\\ 

Color excesses for Cepheids with no literature $E(B-V)$ are estimated
following \cite{2008MNRAS.390.1539M} using mean J-band magnitudes by
\cite{2011ApJS..193...12M}, or single-epoch 2MASS
\citep{2003tmc..book.....C} J-band magnitudes, $m_J(\rm{JD})$.  
The method requires knowledge of the pulsation
period, $P$, and the average J-band magnitude, $\langle m_J \rangle$. If
not known from the literature, the latter can be estimated by
\citep[Eq. 5]{2008MNRAS.390.1539M}:
\begin{equation}
\langle m_J \rangle \simeq m_J({\rm JD}) - \left[ \frac{| m_{V(\phi_J)} -
    \mbox{max}_V |}{\mbox{amp}_V} - 0.5 \right] \cdot 0.37 \mbox{amp}_V \,,
\label{eq:meanJ} 
\end{equation}
where $\phi_J$ denotes the pulsation phase of the J-band measurement, and $m_{V(\phi_J)}$ is the V-band magnitude at that phase. $E(B-V)$ can then be estimated by
\begin{equation}
E(B-V) = -0.270\log{P} + 0.415 \left( \langle m_V \rangle - \langle m_J \rangle
\right ) - 0.255\,.
\label{eq:ebvJ}
\end{equation}
Wherever possible, $m_{V(\phi_J)}$ was obtained from the ASAS light curve. If
this is impossible, we assume a sinusoidal light curve with the given mean
magnitude and (semi-)amplitude.

\begin{figure*}
\centering
\includegraphics{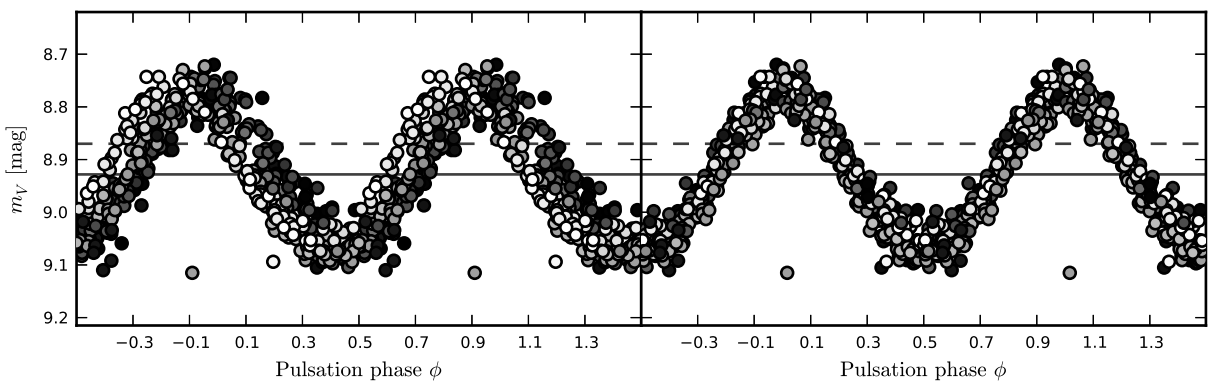}
\caption{ASAS-3 V-band phase-folded light curve of QZ\,Nor. Grayscaled
symbols (from black to white) indicate increasing Julian date of
observation.
\textit{Left panel:} pulsation period and epoch of maximum light from the ACVS.
\textit{Right panel:}  optimized period and epoch used.}
\label{fig:ASASephem}
\end{figure*}
Uncertainties or changes in pulsation period can significantly impact the phase
calculated for the single-epoch 2MASS measurement, $\phi_J$. We therefore
optimize Cepheid ephemerides for which ASAS data were available. To do so, we
compute a grid (at fixed periods) of Fourier series fits around the period
provided in the ACVS and retain the solution with the minimum root mean square.
Epochs are optimized by simply shifting the phase-folded curve.
Figure\,\ref{fig:ASASephem} illustrates this step for the overtone Cepheid
QZ\,Nor. We then take care to employ the most recently determined pulsation
ephemerides available and estimate reddening uncertainties using error
propagation for the quantities involved.

We note that this approach may be subject to multiple issues such as: i) the
unknown shape of the light curve; ii) the applicability of Eq.\,\ref{eq:ebvJ};
iii) period changes that impact $m_{V(\phi_J)}$; iii) the approximate form of
the relationship in Eq.\,\ref{eq:meanJ}. We therefore compare the 2MASS-based
color excesses to the reference values, see Fig.\,\ref{fig:cepebv}, where the
result of a weighted least squares fit is indicated by a straight line and
does not differ much from the diagonal indicated by a dashed line.
Despite considerable dispersion, the correspondence is clear and the results are
promising (RMS of $0.1$\,mag).

\subsubsection{Proper Motions}
Cepheid proper motions are taken from the following sources in order of preference:
\begin{enumerate}
  \item Hipparcos proper motions from the new reduction by \cite{2007ASSL..350.....V}
  \item The PPMXL catalog by \cite{2010AJ....139.2440R}
\end{enumerate}

\subsubsection{Systemic Radial Velocities} \label{sec:CepRV}
\paragraph{New Observations}\label{sec:CepheidRV}
\begin{figure*}
\centering
\includegraphics{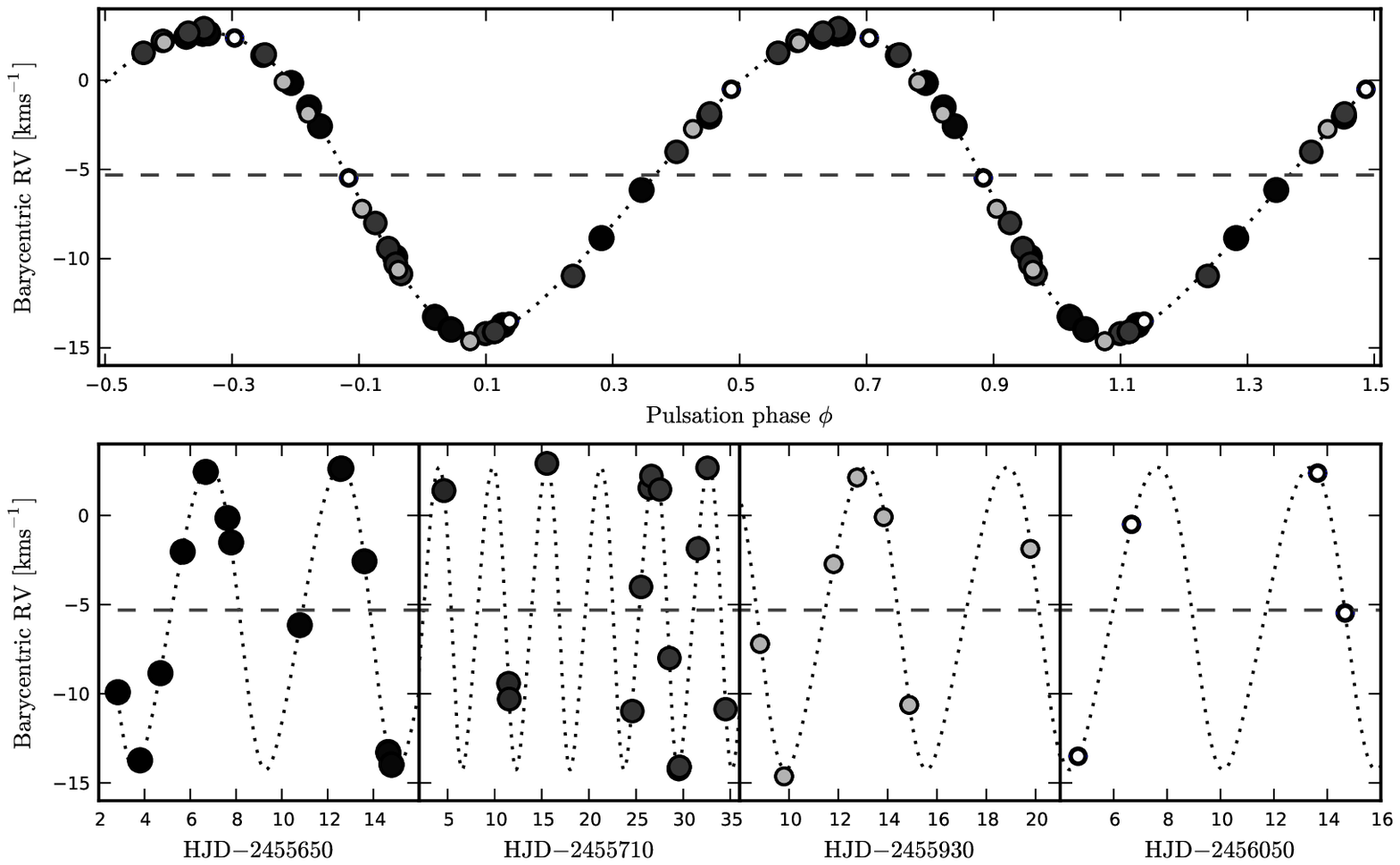}
\caption{RV data recently published in
\citet{2013MNRAS.430.2018S} for first-overtone pulsator GH\,Car, obtained in
southern hemisphere with \textit{CORALIE} as part of our program.
Grayscale (black to white) and size-code (larger to smaller) indicate
increasing Julian Date of observation. Fit of triple-harmonic Fourier series
indicated by a dotted line. $v_\gamma$ indicated by a dashed horizontal line. 
\textit{Top panel}: phase-folded, 
\textit{Bottom panel}: non-folded, showing zooms of the four 
observing runs during which data were obtained. } \label{fig:GHCar} 
\includegraphics{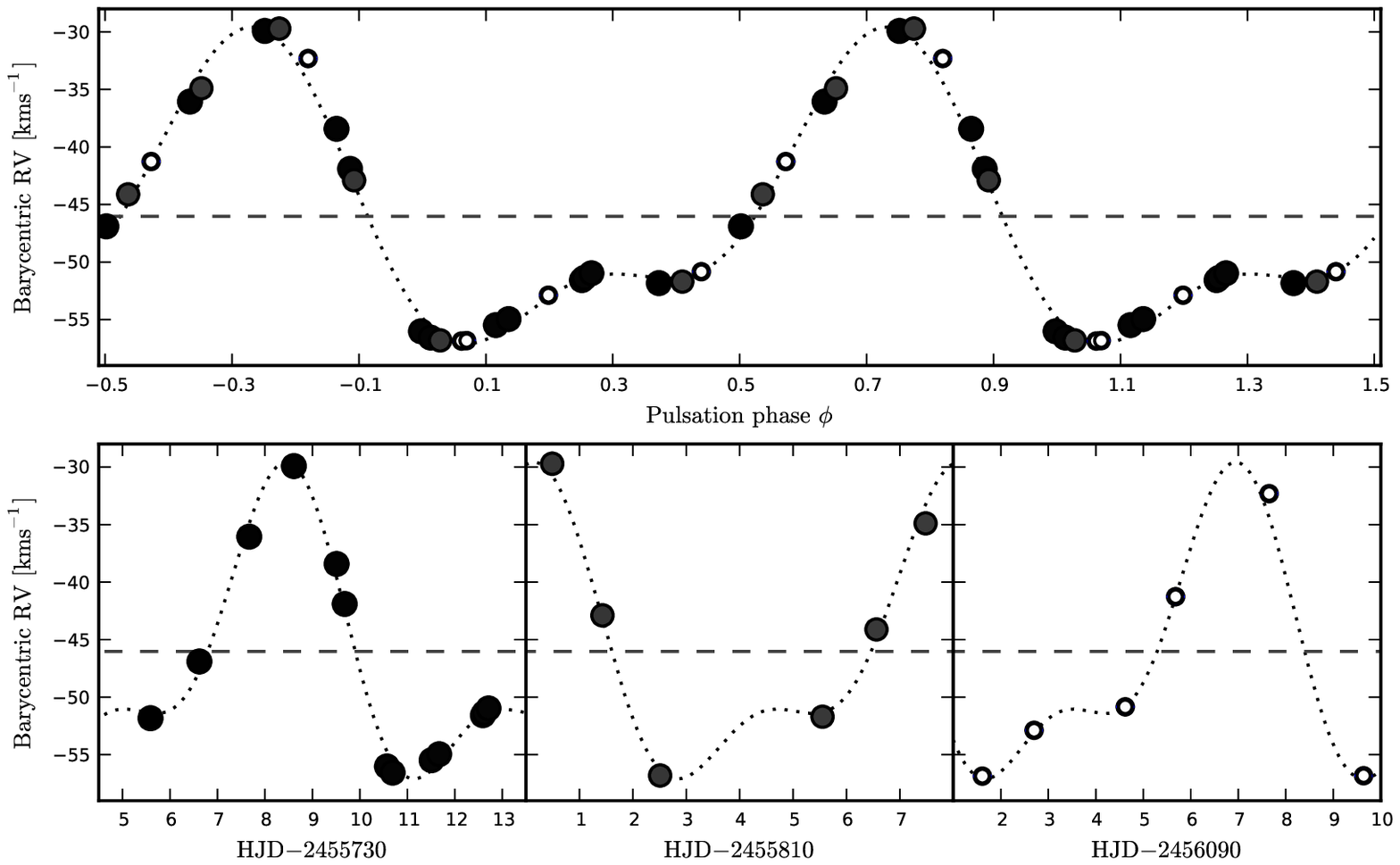}
\caption{New RV data for fundamental-mode Cepheid V$2340$\,Cyg, obtained in
northern hemisphere with \textit{HERMES}. Grayscale (black to white) and
size-code (larger to smaller) indicate increasing Julian Date of observation.
Fit of triple-harmonic Fourier series indicated by dotted line. $v_\gamma$
indicated by dashed horizontal line. \textit{Top panel}: phase-folded,
\textit{Bottom panel}: non-folded, showing zooms of the three observing runs
during which data were obtained.}      
\label{fig:V2340Cyg}
\end{figure*}
In order to extend the number of Cepheids with known
systemic radial velocities, $v_\gamma$, we\footnote{The authors are grateful to the observers who contributed in this effort; names are given in the acknowledgments.} carried
out observations between November 2010 and July 2012 using the fiber-fed
high-resolution echelle spectrographs \textit{CORALIE}
(\citealt{2001Msngr.105....1Q}, see also the instrumental upgrades described in
\citealt{2010A&A...511A..45S}, $R \sim 60000$)  at the $1.2$\,m Euler telescope
at La Silla, Chile, and \textit{HERMES} (\citealt{2011A&A...526A..69R}, $R \sim
80000$)  at the identically-built Mercator telescope on La Palma. In total, we
observed $103$ Cepheids with \textit{CORALIE} and $63$ with \textit{HERMES}.
$18$ Cepheids were observed with both instruments, i.e. from both hemispheres.
For $85$ of these Cepheids, no radial velocity (RV) data are available in the literature.

Efficient reduction pipelines exist for both instruments that include
pre- and overscan bias correction, cosmic removal, as well as flatfielding
using Halogen lamps and background modelization. ThAr lamps are used for the
wavelength calibration.

The RVs are computed via the cross-correlation technique described
in \cite{1996A&AS..119..373B}. We use numerical masks designed for solar-like
stars (optimized for spectral type G2) for all cross-correlations. Both
instruments are very stable and yield very high precision RVs of $\sim 10$\,ms$^{-1}$ 
\citep{2000A&A...354...99Q,2011A&A...526A..69R}. The measurement uncertainty
is therefore not limited by the instrumental precision, but by line asymmetries due to  pulsation. A detailed investigation of these effects is out of scope for this paper and will be presented in a future publication. The typical uncertainty on individual measurements is thus at the $100-300$\,ms$^{-1}$ level, depending on the star and pulsation phase.

\paragraph{Literature Data}\label{sec:RVlit}
In addition to the previously unpublished radial velocities described in Sec.\,\ref{sec:CepheidRV}, we employ literature data from many references to determine systemic velocities, $v_\gamma$, see Sec.\,\ref{sec:vgamma}. 
The addition of literature RVs extends the baseline of our otherwise relatively short (1.5 years) observing program, thereby enhancing our sensitivity to binarity. 
For binary
Cepheids\footnote{cf. L. Szabados' database of binary Cepheids available at
\texttt{http://www.konkoly.hu/CEP/nagytab3.html}} 
 with published orbital solutions, we adopt the literature $v_\gamma$.

Aside from the systemic RVs in the Fernie database and KS$09$, we compile RV
time series from the following sources:
\cite{1980SAAOC...1..257L,1981ApJS...46..287G,1985ApJ...295..507G,1983ApJ...272..214E,1985ApJS...57..595C,1985SAAOC...9....5C,1987ApJS...65..307B,1988ApJS...66...43B,1989AJ.....98.1672G,1989ApJS...69..951W,1991ApJS...76..803M,1992AJ....103..529M,1998AJ....115..635M,1992SvAL...18..316G,1996AstL...22..175G,1998AstL...24..815G,2002yCat.3229....0G,1993PASP..105..836E,1994A&AS..105..165P,1997A&A...318..416P,1994A&AS..108...25B,2002ApJS..140..465B,1998MNRAS.297..825K,1999A&AS..140...79I,2004A&A...415..531S,2005ApJS..156..227B,2005MNRAS.362.1167P,2009MNRAS.396.2194B}.
The data for most of these sources dated earlier than $1986$ are extracted from
the McMaster Cepheids database. Newer data are obtained
through VizieR\footnote{\texttt{http://vizier.u-strasbg.fr/viz-bin/VizieR}}.

\paragraph{Systemic Radial Velocities, $\mathbf{v_\gamma}$}\label{sec:vgamma} The systemic radial velocity, $v_\gamma$, is
obtained by fitting a Fourier series to the RVs. We use pulsation period, $P$, as a fixed parameter, since it is known for all Cepheids we observed. 

The basic analytical form applied was a Fourier series with $n$ harmonics and phase $\phi$ is:
\begin{equation}
  FS_n = v_{\gamma} + \sum_{n=1,2,3,...}{a_{n} \sin{(2 n \pi \phi)} + b_{n}
  \cos{(2 n \pi \phi)}}
  \label{eq:Fseries}
\end{equation}

Since the number of data points available varies for each star, we do not fix the number of harmonics in this fit. Instead, we
iteratively increase the degree of the Fourier series until an F-test indicates an overly 
complex representation, i.e. when spurious fit improvement is more likely than
$0.27\,\%$.  For some stars, we therefore use only a simple sine
function, whereas stars with many measurements are fitted using up to five harmonics. We show two examples of newly observed RV curves in Figs.\,\ref{fig:GHCar} and \ref{fig:V2340Cyg}. A full description and publication of the new radial velocity data will follow in the near future.

We adopt a fixed error budget of $3$\,km\,s$^{-1}$ on $v_\gamma$. Although this may overestimate the uncertainties for some very good cases, it is intended to account for a range of systematic errors, such as unseen binarity, instrumental zero point differences and insufficient phase coverage. 
We are confident that this error budget is sufficiently large to prevent the exclusion of good member candidates, while being sufficiently small to provide a stringent constraint. If insufficient data points render the Fourier fit unsatisfactory, we determine a rough estimate of $v_\gamma$ and its error budget by eye.

\subsubsection{Iron Abundance}
We rely mostly on iron abundances by \cite{2011AJ....142..136L} and
complement these with the compilation in KS$09$ that made the enormous effort of
homogenizing iron abundances from the literature available, namely
from
\cite{1983JApA....4...75G,1997AJ....113.1073F,2002A&A...381...32A,2002A&A...384..140A,2002A&A...392..491A,
2003A&A...401..939L,2004A&A...420..655G,2004A&A...413..159A,2005AJ....130.1880A,
2005AJ....129..433K,2005PASP..117.1173K,2005A&A...429L..37R,Mottini2006PhD,2006AJ....131.2256Y,2007A&A...467..283L},
as well as the work by
\cite{2007A&A...473..579S,2008A&A...490..613L,2008A&A...488..731R}.

Unfortunately, the adopted standard value of solar iron abundance can vary among
references, possibly introducing systematic offsets between different authors'
studies.  Furthermore, an estimation of the iron abundance in a Cepheid is more
complex than in a non-pulsating star, since the stellar parameters (e.g.
temperature and turbulence) vary during the pulsation cycle, and since the
atmosphere is not static. We therefore adopt generous error budgets  of
$0.1$\,dex for the values from \cite{2011AJ....142..136L}, and $0.15$\,dex in
[Fe/H] for the others. 

\subsubsection{Age}\label{sec:CepAges}
Cepheid ages can be calculated for first overtone and
fundamental mode pulsators using the period-age (PA) relations given in
\cite{2005ApJ...621..966B}. For fundamental mode Cepheids, we use $\log{t} = \left(8.31 \pm 0.08\right) - \left(0.67 \pm 0.01\right)\log{P}$. For overtone pulsators, the relation used 
is $\log{t} = \left(8.08 \pm 0.04\right) - \left(0.39 \pm 0.04\right)\log{P}$. 
Age error budgets are calculated from the uncertainties stated for slope and intercept.

\section{Results}\label{sec:Results}

This section presents the results from our census. As
mentioned in Sec.\,\ref{sec:OCcompilation}, some host Cepheid clusters known in the literature
\citep[e.g. in][or T10]{2002AJ....124.2931T} are not present in our cluster sample.
Such cases are briefly mentioned in Sec.\,\ref{sec:missedCCs}. We furthermore
note that stellar associations are not considered here.

In addition to the membership probabilities computed,
we consider the quality of the data employed to constrain membership, and
compare our results to the published literature.
We then flag Combos as \textit{bona fide}, inconclusive, unlikely, or
non-members. These flags are attributed according to the following
reasoning:
\begin{itemize}
  \item \textit{bona fide} is attributed to Combos with typically high
  priors and high likelihoods constrained by multiple parameters, in particular parallax. Closer
  inspection of the individual membership employed or the literature builds
  confidence in membership. Some Combos studied in detail in the literature
  prove to be strong candidates, despite low probabilities computed here,
  pointing to limitations of the data used as input in our analysis. We consider
  these Combos \textit{bona fide} members.
  \item \textit{inconclusive} CCs are candidates for which the
  membership constraints available are insufficient to consider them
  \textit{bona fide}, e.g. if $P(A) > 0.5$ with no additional membership
  constraints. We flag newly-identified Combos as
  \textit{inconclusive}, if the prior vanishes and $ 0.1 < P(B|A) < 0.8$ has
  been computed from at least 3 membership constraints that exceed the
  combined error budgets. These candidates warrant follow-up.
  \item \textit{unlikely} CCs have low likelihoods ($< 10\%$) due to discrepant membership constraints (more than one constraint off by
  approximately $2\sigma$), although evidence supporting membership may exist in
  the literature. Membership cannot be ruled out altogether for these
  candidates that may benefit from additional follow-up. 
  \item \textit{non-members} form the majority of Combos cross-matched. They
  are clearly inconsistent with membership.
\end{itemize}

It should be kept in
mind that our analysis is of a statistical nature may not provide the final answer for every Combo.
While the benefit of our analysis is a consistent and transparent approach to
determining membership, the correctness of our membership probabilities relies
entirely on the accuracy of the input data; this is particularly true for
reddening and distances, or pulsation modes. Therefore, we caution that CCs
previously discussed in the literature that are found to be unlikely or
non-members by our analysis should not be discarded fully without additional
consideration or follow-up.

We start the presentation of our results with CCs known from the
literature (Sec.\,\ref{sec:litCombos}). While we take care to include relevant
references, it is almost inevitable that some works are overlooked in a field
with this much history. The literature CCs are followed by new
candidates and other newly-identified Combos of interest
(Sec.\,\ref{sec:newCombos}). For brevity of the main body, we defer
presentation of inconclusive, unlikely, and inconsistent Combos to appendix A.  

In Sec.\,\ref{sec:GalPLR}, we then revisit the Galactic
Cepheid period-luminosity relationship using our \textit{bona fide} CC sample.

\subsection{Literature Combos}\label{sec:litCombos}
The main references considered for CCs are \cite{1999PASP..111..775F},
Turner \& Burke 2002, and \citet[from
hereon: T$10$]{2010Ap&SS.326..219T}. Additional Cepheids whose cluster
membership was considered in the literature are mentioned where appropriate, cf.
also the references given in the caption of Tab.\,\ref{tab:literature} and
Sec.\,\ref{sec:missedCCs}.

\begin{table*} 
 \centering 
 \caption{Results of our membership analysis for combinations known in the
 literature. Constraints indicated are parallax, $\varpi$, radial velocity,
 $v_r$, proper motion, $\mu_\alpha^*$ and $\mu_\delta$, iron abundance [Fe/H],
 and age. 
 Filled circles indicate constraints consistent between Cepheid and cluster, 
 greater deviations are stated explicitly in units of the square-summed
 uncertainties. Open circles indicate unavailable membership constraints. 
$R_{\rm{Cl}}$ denotes the distance in parsecs of the
 Cepheid from cluster center, assuming membership and the cluster's heliocentric
 distance. $P(A)$ is the prior used, asterisks mark $P(A)$ values based on D02
 apparent diameters. 
 Column $P(B|A)$ lists likelihoods, and $P(A|B)$ the combined membership probability. The last column $\rm{CC}$ indicates qualitatively, how membership is judged for a particular  Combo: `y' denotes \textit{bona fide} CCs; `i' denotes that the data available
 yield an inconclusive result; `u' denotes unlikely membership; `n' denotes 
 Combos that are clearly inconsistent with membership. Column Ref. lists
 some pertinent references:
  	a: \citet{1955MNSSA..14...38I}, 
  	b: \citet{1956PZ.....11..325K}, 
  	c: \citet{1957MNRAS.117..193F}, 
  	d: \citet{1958ApJ...128..150S}, 
  	e: \citet{1980IBVS.1853....1E}, 
  	f: \citet{1982PASP...94.1003T}, 
  	g: \citet{1985MNRAS.214...45W}, 
  	h: \citet{1986AJ.....92..111T}, 
  	i: \citet{1992AJ....104.1132T}, 
  	j: \citet{1995AJ....110.2280M}, 
  	k: \citet{1997AJ....113.2104T}, 
  	l: \citet{2003MNRAS.345..269H}, 
   	m: \citet{2007ApJ...671.1640A},
   	n: \citet{2010Ap&SS.326..219T}, 
   	o: \citet{1981AJ.....86..231T},
   	p: \citet{1978ApJ...224..948F}, 
   	q: \citet{1985MNRAS.213..889W},
   	r: \citet{1998AJ....115.1958T}, 
   	s: \citet{1985AJ.....90.1231T},
   	t: \citet{1980ApJ...240..137T}, 
   	u: \citet{1966ZA.....64...54Y},
   	v: \citet{1998AJ....116..274T}, 
   	w: \citet{2002AJ....124.2931T},
   	x: \citet{1976AJ.....81.1125T}, 
   	y: \citet{1987MNRAS.229...31W},
   	z: \citet{1994AJ....107.1796T},
   	A: \citet{1993ApJS...85..119T},
   	B: \citet{1992AJ....104.1865T}, 
   	C: \citet{2011ApJ...741L..27M},
   	D: \citet{1990A&AS...86..209V},
   	E: \citet{2000A&AS..146..251B},
   	F: \citet{2008MNRAS.388..444T},
   	G: \citet{1995MNRAS.277..250B},
   	H: \citet{1977PASP...89..277T} } 
\begin{tabular}{@{}llccccccllllrr@{}} 
\hline 
Cluster 	 & Cepheid 	 & \multicolumn{6}{|c|}{Constraints} 	 & R$_{\rm cl}$ 	 & $P(A)$ 	 & $P(B \vert A)$ 	 & $P(A \vert B)$ & CC & Ref. \\ 
 & & $\varpi$ & $v_r$ & $\mu_\alpha^*$ & $\mu_\delta$ & [Fe/H] & age & [pc] &  & 
 & & & \\
\hline 
IC 4725	 & U Sgr	 & $\bullet$ & $\bullet$ & $\bullet$ & $\bullet$ & $\bullet$ & $\bullet$	 & 0.3	 & 1.0	 & 0.984	 & 0.984  & y & abcm \\  
NGC 7790	 & CF Cas	 & $\bullet$ & $\bullet$ & $\bullet$ & $\bullet$ & $\circ$ & $\bullet$	 & 0.9	 & 0.955	 & 0.975	 & 0.931 &  y & dj   \\  
NGC 129	 & DL Cas	 & $\bullet$ & $\bullet$ & $\bullet$ & $\bullet$ & $\circ$ & $\bullet$	 & 0.2	 & 1.0	 & 0.857	 & 0.857 &  y & bi   \\  
Turner 9	 & SU Cyg	 & $\bullet$ & $\bullet$ & $\bullet$ & $\bullet$ & $\circ$ & $1.2\sigma$	 & 0.0	 & 1.0	 & 0.807	 & 0.807 &  y & kn  \\  
NGC 7790	 & CE Cas A	 & $\bullet$ & $\bullet$ & $\bullet$ & $\bullet$ & $\circ$ & $\bullet$	 & 1.5	 & 0.71	 & 0.975	 & 0.693 &  y & dj  \\  
NGC 7790	 & CE Cas B	 & $\bullet$ & $\bullet$ & $\bullet$ & $\bullet$ & $\circ$ & $\bullet$	 & 1.6	 & 0.697	 & 0.956	 & 0.666 &  y & dj \\  
NGC 6649	 & V367 Sct	 & $\circ$ & $\bullet$ & $\bullet$ & $1.3\sigma$ & $\circ$ & $\circ$	 & 1.0	 & 0.884$^{*}$	 & 0.65	 & 0.574 &  y & op  \\  
NGC 6067	 & V340 Nor	 & $\bullet$ & $\bullet$ & $\bullet$ & $\bullet$ & $\bullet$ & $1.9\sigma$	 & 0.6	 & 1.0	 & 0.573	 & 0.573 &  y & gmnl \\  
Lyng\aa\ 6	 & TW Nor	 & $1.2\sigma$ & $\bullet$ & $\bullet$ & $1.3\sigma$ & $\circ$ & $\bullet$	 & 0.6	 & 1.0$^{*}$	 & 0.453	 & 0.453 &  y & nmqC  \\  
vdBergh 1	 & CV Mon	 & $\bullet$ & $\bullet$ & $2.2\sigma$ & $\bullet$ & $\circ$ & $\bullet$	 & 0.6	 & 1.0	 & 0.318	 & 0.318 &  y & r  \\  
NGC 6087	 & S Nor	 & $\bullet$ & $\bullet$ & $1.2\sigma$ & $2.0\sigma$ & $1.2\sigma$ & $1.3\sigma$	 & 0.6	 & 1.0	 & 0.192	 & 0.192 &  y & abch \\  
Trumpler 35	 & RU Sct	 & $1.3\sigma$ & $\bullet$ & $\bullet$ & $\bullet$ & $\circ$ & $\bullet$	 & 5.1	 & 0.194$^{*}$	 & 0.840	 & 0.163  &  y & nltu  \\
Collinder 394	 & BB Sgr	 & $1.0\sigma$ & $\bullet$ & $\bullet$ & $1.2\sigma$ & $\circ$ & $\bullet$	 & 3.7	 & 0.208	 & 0.637	 & 0.133  &  y & ns  \\  
Turner 2	 & WZ Sgr	 & $\bullet$ & $\circ$ & $\bullet$ & $2.2\sigma$ & $\circ$ & $\bullet$	 & 5.3	 & 0.337$^{*}$	 & 0.287	 & 0.097  & y & A  \\
Trumpler 18	 & GH Car	 & $\circ$ & $1.4\sigma$ & $\bullet$ & $\bullet$ & $\circ$ & $2.0\sigma$	 & 2.9	 & 0.194	 & 0.143	 & 0.028 & u &  DE  \\  
NGC 6067	 & QZ Nor	 & $\bullet$ & $\bullet$ & $\bullet$ & $\bullet$ & $\bullet$ & $\bullet$	 & 7.4	 & 0.029	 & 0.963	 & 0.027 &  y & eg  \\  
Berkeley 58	 & CG Cas	 & $\bullet$ & $\circ$ & $\bullet$ & $3.3\sigma$ & $\circ$ & $\bullet$	 & 5.0	 & 0.308	 & 0.027	 & 0.008   & y & F  \\  
NGC 5662	 & V Cen	 & $\bullet$ & $\bullet$ & $\bullet$ & $\bullet$ & $\bullet$ & $\bullet$	 & 5.8	 & 0.006	 & 0.958	 & 0.006   &  y & fmn\\  
NGC 6664	 & EV Sct	 & $\bullet$ & $\bullet$ & $\bullet$ & $\bullet$ & $\circ$ & $\bullet$	 & 7.5	 & 0.0	 & 0.866	 & 0.0 & y & wx  \\
Ruprecht 173	 & X Cyg	 & $\circ$ & $\circ$ & $\circ$ & $\circ$ & $\circ$ & $\circ$	 & --	 & 0.878$^{*}$	 & 1.0	 & 0.878 & i & nvw  \\
Dolidze 45	 & V1334 Cyg	 & $\circ$ & $\circ$ & $\circ$ & $\circ$ & $\circ$ & $\circ$	 & -- 	 & 0.017$^{*}$	 & 1.0	 & 0.017 & i & n  \\
\hline 
Ruprecht 79	 & CS Vel	 & $1.8\sigma$ & $1.5\sigma$ & $1.2\sigma$ & $1.9\sigma$ & $\circ$ & $2.3\sigma$	 & 1.5	 & 1.0	 & 0.007	 & 0.007 & i & ny  \\  
Platais 1	 & V1726 Cyg	 & $\circ$ & $3.2\sigma$ & $\bullet$ & $1.0\sigma$ & $\circ$ & $1.8\sigma$	 & 1.4	 & 0.98	 & 0.006	 & 0.006 & i & z  \\  
NGC 1647	 & SZ Tau	 & $\bullet$ & $1.8\sigma$ & $\bullet$ & $2.6\sigma$ & $\circ$ & $1.2\sigma$	 & 20.1	 & 0.0	 & 0.047	 & 0.0  & u & B  \\  
NGC 3496	 & V442 Car	 & $\circ$ & $\circ$ & $1.1\sigma$ & $2.3\sigma$ & $\circ$ & $\circ$	 & 1.0	 & 0.625	 & 0.039	 & 0.024  & n &  G  \\  
King 4	 & UY Per	 & $1.3\sigma$ & $\circ$ & $1.6\sigma$ & $2.6\sigma$ &
$\circ$ & $\bullet$	 & 12.6	 & 0.021 & 0.019	 & 0.0 & n & H  \\
Turner 5	 & T Ant	 & $4.3\sigma$ & $\bullet$ & $3.3\sigma$ & $4.5\sigma$ & $\circ$ & $3.2\sigma$	 & 0.0	 & 1.0	 & 0.0	 & 0.0 & n & w   \\  
NGC 4349	 & R Cru	 & $6.8\sigma$ & $\bullet$ & $2.1\sigma$ & $1.1\sigma$ & $2.1\sigma$ & $2.3\sigma$	 & 9.5	 & 0.048	 & 0.0	 & 0.0 & n & w  \\  
NGC 4349	 & T Cru	 & $6.7\sigma$ & $\bullet$ & $2.1\sigma$ & $\bullet$ & $2.2\sigma$ & $2.5\sigma$	 & 19.9	 & 0.0	 & 0.0	 & 0.0 & n & w  \\  
NGC 2345	 & TV CMa	 & $\bullet$ & $5.6\sigma$ & $\bullet$ & $\bullet$ & $\circ$ & $\bullet$	 & 25.0	 & 0.001	 & 0.0	 & 0.0 & n & xw  \\  
\hline 
\end{tabular}
\label{tab:literature} 
\end{table*}

Table\,\ref{tab:literature} lists the CCs previously discussed in the
literature that are recovered by our analysis. A horizontal line divides cases that we find
can be consistent with membership according to the data compiled (above), and
those that tend to be inconsistent with membership in our analysis (below).
Two essentially unconstrained Combos known in the literature are included above
the horizontal line. 
For each deviant membership constraint, i, we list the level of disagreement
between cluster and Cepheid value, i.e., $\vert x_{\rm i} \vert$
from Eq.\,\ref{eq:xvec}, in units of the square-summed uncertainties
$\sigma_{\rm i}^2 = \sigma_{\rm{Cl,i}}^{2} + \sigma_{\rm{Cep,i}}^{2}  $.

\subsubsection{Missed Combos}\label{sec:missedCCs}
Our analysis is limited to open clusters listed in D02. The following
CCs reside in nearby sparse clusters that are not included in D02 and
could thus not be studied by our analysis:
$\alpha$\,UMi \citep[][but see also
\citealt{2013A&A...550L...3V}]{2013ApJ...762L...8T};
$\delta$\,Cep \citep[e.g.][]{2012ApJ...747..145M}; 
$\zeta$\,Gem \citep{2012ApJ...748L...9M}; 
SU\,Cas \citep{2012ApJ...753..144M,2012MNRAS.421.1040M,2012MNRAS.422.2501T}.

\subsubsection{Bona-fide CCs}
Based on the available data and literature, we flag the following
literature Combos as \textit{bona fide} CCs, cf. Tab.\,\ref{tab:literature}: 
U\,Sgr in IC\,4725; CF\,Cas, CE Cas A \& B in NGC\,7790; DL\,Cas in NGC\,129; 
SU\,Cyg in Turner\,9; V367\,Sct in NGC\,6649, V340\,Nor and QZ\,Nor in
NGC\,6067; TW\,Nor in Lyng\aa\ 6; CV\,Mon in vdBergh 1; S\,Nor in NGC\,6087; BB\,Sgr in
Collinder\,394; RU\,Sct in Collinder\,394; CG\,Cas in Berkeley\,58; V\,Cen in
NGC\,5662. For more information on those Combos with high priors and
high likelihoods, i.e., the more or less obvious members, we
refer to the original references listed in Tab.\,\ref{tab:literature}, 
as well as to the data table supplied in electronic form.
As mentioned in Sec.\,\ref{sec:Results}, some Combos flagged as
\textit{bona fide} CCs require inspection of the available data and literature
in addition to the membership probabilities in order to conclude on membership.
We discuss these combos in the paragraphs below.

\paragraph{V340\,Nor and QZ\,Nor in NGC 6067}
We find two Cepheids that appear to belong to NGC\,6067, 
namely V340\,Nor, which lies within its core radius, and QZ\,Nor, an overtone
pulsator (KS09) that lies outside $r_{\rm lim}$. 
Cluster membership for QZ\,Nor was  first considered by \cite{1983AJ.....88..379E}, and by
\cite{1985MNRAS.214...45W} for V340\,Nor. All membership constraints were
employed for both Cepheids, and both are  consistent with membership for the
open cluster data listed in D02. The only discrepant constraint is age for
V340\,Nor.

If both Cepheids belong to the same cluster, then their respective
membership constraints should agree. Interestingly, the distance estimate of QZ\,Nor by
\cite{2011A&A...534A..94S} is much closer to the cluster's, while there are
nearly $400$pc difference between the estimates for both Cepheids. In terms of
parallax, QZ\,Nor ($0.74 \pm 0.07$ [mas]) is consistent with NGC\,6067 ($0.71
\pm 0.14$ [mas]), but slightly off from V340\,Nor ($0.58 \pm 0.09$
[mas]). However, the difference in $v_\gamma$ between the two Cepheids is
minimal ($0.73\,\rm{km\,s^{-1}}$). Given that V340\,Nor is a visual binary, the
small offset in proper motion between the two Cepheids is not alarming, and
[Fe/H] is indistinguishable. In terms of age, QZ\,Nor seems to be
slightly older than V340\,Nor ($7.85\pm0.07$ vs. $7.60\pm0.08$), and better
matches to the cluster's age ($8.08\pm0.23$, D02).

In summary, the two Cepheids have differing parallax and age. The
cluster values from D02 happen to lie between the two, oddly enough favoring
QZ\,Nor, which lies at greater separation from the cluster's core. Therefore,
the cluster parameters may require reconsideration. Since NGC\,6067 is located in
the Norma cloud (cf. 
atlas page in K$05$), the determination of cluster radii is rather difficult.
Differential reddening may be important to resolve this conundrum (higher for
V340\,Nor which is closer to cluster center). We therefore note that there are
some issues with the membership constraints employed here, and detailed
follow-up of the cluster is required. Until then, the constraints compiled are
consistent with membership for both Cepheids, and we consider both to be 
bona fide cluster members.

\paragraph{CV\,Mon and van\,den\,Bergh\,1} \label{sec:cvmon} 
CV\,Mon lies right in the center of cluster van den Bergh\,1 and was
studied in detail by \cite{1998AJ....115.1958T}. 
The constraints employed are parallax, radial velocity,
proper motion, and age, and yield a likelihood of $32\%$, which is low due to  
the discrepant $\mu_\alpha^*$. The average cluster RV determined by
\cite{1999AstL...25..595R} is identical to that of CV\,Mon and no information on
the number of stars involved in its determination is given; it should thus be
discarded as membership constraint (this would lower the likelihood to $20\%$).
Aside from this, the cluster data from D$02$ is largely consistent with the data
for the Cepheid. Parallax (also reddening), $\mu_\delta$, and age agree well
between cluster and Cepheid. Thus, the Cepheid likely lies inside the
volume occupied by the Cluster, and therefore should be considered to be a bona
fide CC. Observational follow-up of cluster proper motion and radial velocity
would be beneficial.

\paragraph{S\,Nor and NGC\,6087} \label{sec:snor} S\,Nor's membership in
NGC\,6087 was among the first to ever be suggested \citep{1955MNSSA..14...38I}
and confirmed using radial velocities \citep{1957MNRAS.117..193F}, as well as
the detailed study by \cite{1986AJ.....92..111T} based on reddening and
distance.

The values for the cluster's mean RV differ greatly between D$02$
(6\,km\,s$^{-1}$), K$05$ ($-9$\,km\,s$^{-1}$), and
\citet[$2.0$\,km\,s$^{-1}$]{1957MNRAS.117..193F}, which is important considering
the Cepheid's $v_\gamma = 2.53$\,km\,s$^{-1}$ \citep{2008A&A...488...25G}. We
note that the value adopted by D$02$ is measured on the Cepheid itself
\citep{2008A&A...485..303M}, although without taking orbital motion into
account, and is therefore not suitable as a membership constraint. Since
\cite{1957MNRAS.117..193F} investigated the largest number of stars and
specifically targeted this cluster, we trust that their $2.0$\,km\,s$^{-1}$ is
the best available estimate for the mean velocity of the cluster.

We note that proper motion, reddening, metallicity, and age are
slightly discrepant between cluster and Cepheid, resulting in a low likelihood.
However, these differences barely exceed the combined uncertainties and may not
be significant. 

\paragraph{RU\,Sct and Trumpler\,35} 
Based on age and distance from D02, this Combo
would appear nearly inconsistent with membership. Unfortunately, the average
cluster RV appears to have been measured on the Cepheid
\citep{1999AstL...25..595R} and can therefore not be considered a
valid membership constraint.
Membership of RU\,Sct in Trumpler\,35 was studied in
detail by \cite{1980ApJ...240..137T},
\cite{2003MNRAS.345..269H}, and T10.
Closer inspection of these references reveals an underestimated cluster distance
in D02 \citep[compared also to][]{1966ZA.....64...54Y}. The region around
the Cepheid contains multiple associations, which may explain the confusion in D02. 
Using the cluster parallax and age from \cite{1980ApJ...240..137T}, we compute a
likelihood of $84\,\%$, and consider RU\,Sct a
\textit{bona fide} member of Trumpler\,35.

\paragraph{BB\,Sgr and Collinder\,394} BB\,Sgr lies at $21'$ separation
from Collinder\,394's center, i.e., at a distance of $3.7$\,pc assuming membership. 
This Combos was first studied in detail by \cite{1985AJ.....90.1231T} after
having been originally suggested by \cite{1966ATsir.367....1T}. Most membership
constraints could be employed and there are only small discrepancies in parallax
and $\mu_\delta$. The low prior may be misleading in this case, since the
high likelihood indicates membership. We thus consider BB\,Sgr a \textit{bona
fide} member of Collinder\,394.

\paragraph{WZ\,Sgr and Turner\,2}\label{sec:WZSgr} This Combo was first discussed by
\cite{1993ApJS...85..119T} when the cluster was first discovered. Most
membership constraints compiled from D02, i.e., parallax, $\mu_\delta$, and age,
differ between Cepheid and cluster, which would result in a likelihood 
of $<1\,\%$.
The cluster parameters listed in D02 were taken from the automated, 2MASS-based
study by \cite{2008NewA...13..370T}. However, the much more detailed study by
\cite{1993ApJS...85..119T} should be given higher weight, especially
for its thorough treatment of reddening, and the more precise photometry used.
Hence, we compute the likelihood using the cluster parameters for parallax and age from \cite{1993ApJS...85..119T} and find a combined membership
probability of $10\,\%$.
The sole discrepant membership constraint remains proper motion. However, this
discrepancy alone is not sufficiently strong to indicate non-membership.

\paragraph{CG\,Cas - Berkeley\,58}
CG\,Cas lies at a separation of $5.7'$ (outside $r_{\rm c}$) from
Berkeley\,58's center, at roughly half the limiting radius. While the
Cepheid's PLR-based parallax is close to that of the cluster and reddening is
in agreement, $\mu_\delta$ is discrepant by $3.3\,\sigma$ and does not suggest a
common point of origin. However, \cite{2008MNRAS.388..444T}
conclude in favor of membership based on a detailed study involving age,
reddening, distance and radial velocity.
Since the cluster is located in  the Perseus spiral arm, the
proper motion estimate in K12 may well be dominated by Galactic motion. Hence,
the likelihood computed here is likely underestimated, and we should trust the
result by \cite{2008MNRAS.388..444T}.

\paragraph{V\,Cen and NGC\,5662} Despite the low prior,
all membership constraints indicate V\,Cen's membership in NGC\,5662, yielding a very
high likelihood of $92\,\%$. At NGC\,5662's distance of $666$\,pc, the
Cepheid lies $5.8$\,pc from cluster center. Hence, the prior may be misleading
in this case, perhaps due to underestimated radii. We therefore consider V\,Cen a
\textit{bona fide} CC of NGC\,5662\footnote{According to Turner (2013, priv.
comm.), NGC\,5662 is actually a double cluster, and V\,Cen belongs to
NGC\,5662b.}.

\paragraph{EV\,Sct and NGC\,6664} This Combo, mentioned previously in
\cite{1976AJ.....81.1125T} and \cite{2002AJ....124.2931T}, would be nearly
inconsistent with membership if the cluster values listed in D02 are employed in the
calculation: the highly discrepant age (2.5$\sigma$) and parallax ($0.57 \pm
0.09$ vs. $0.86 \pm 0.17$) would result in a likelihood
of $13\,\%$.
However, a literature study reveals that the distance and age listed in D02 
may be wrong.
The distance by \citep{1958ApJ...128..166A} and \cite{1982AJ.....87.1197S} are
both much greater than the 1.1\,kpc in D02. We thus adopt the distance and age
by \cite[$1.4$\,kpc]{1982AJ.....87.1197S} and obtain a very high likelihood of
$87\,\%$, using also proper motion and radial velocity. As already noted by
\cite{2007MNRAS.377..147L}, a modern follow-up campaign is warranted for this cluster.

\subsection{New Combos of Interest}\label{sec:newCombos}
Our results suggest the following new \textit{bona fide} CCs: 
SX\,Car in ASCC\,61, ASAS\,J182714-1507.1 in Kharchenko\,3, S\,Mus in ASCC\,69,
UW\,Car in Collinder\,220, and V379\,Cas in NGC\,129, see
Tab.\,\ref{tab:newCombos}. We discuss these Combos in some detail in the
paragraphs below, followed by the identification of some 
unconstrained high-prior Combos recovered by our work. Some Combos flagged as
inconclusive or unlikely members are discussed in appendix A.

\begin{table*} 
 \centering 
 \caption{New Combos of interest. Columns are described in
 Tab.\,\ref{tab:literature}.
 We visually separate a) Combos with membership constraints that lie inside the core of a cluster, b) Combos with high likelihood for which $\varpi$ was available, c) Combos without $\varpi$ that yield high likelihoods, and d) Combos for which no likelihoods could be computed, but that lie close to the core of their potential host clusters. Combos judged unlikely (`u') or bona-fide (`y') are discussed separately in the text. Inconclusive Combos (`i') require additional data or stronger membership constraints.}
\begin{tabular}{@{}llccccccllllr@{}} 
\hline 
Cluster 	 & Cepheid 	 & \multicolumn{6}{|c|}{Constraints} 	 & R$_{\rm cl}$ 	 &
$P(A)$ 	 & $P(B \vert A)$ 	 & $P(A \vert B)$ & CC \\
 & & $\varpi$ & $v_r$ & $\mu_\alpha^*$ & $\mu_\delta$ & [Fe/H] & age & [pc] &  & 
 & &
 \\
\hline 
ASCC 60	 & Y Car	 & $\circ$ & $\bullet$ & $\bullet$ & $\bullet$ & $\circ$ & $\circ$	 & 0.3	 & 1.0	 & 0.786	 & 0.786 & i  \\  
\hline
ASCC 61	 & SX Car	 & $\bullet$ & $\circ$ & $\bullet$ & $\bullet$ & $\circ$ & $\bullet$	 & 20.4	 & 0.001	 & 0.919	 & 0.001  & y \\  
Kharchenko 3	 & ASAS\,J182714-1507.1	 & $\bullet$ & $\circ$ & $\bullet$ &
$\bullet$ & $\circ$ & $\bullet$	 & 43.9	 & 0.004$^{*}$	 & 0.905	 & 0.004  & y   \\
ASCC 69	 & S Mus	 & $\bullet$ & $\bullet$ & $\bullet$ & $\bullet$ & $\circ$ & $1.0\sigma$	 & 11.4	 & 0.004	 & 0.879	 & 0.004  & y  \\  
Collinder 220	 & UW Car	 & $\bullet$ & $\bullet$ & $\bullet$ & $\bullet$ & $\circ$ & $1.1\sigma$	 & 47.2	 & 0.001$^{*}$	 & 0.838	 & 0.001  & y  \\  
IC 4725	 & Y Sgr	 & $1.2\sigma$ & $\bullet$ & $\bullet$ & $\bullet$ & $\bullet$ & $\bullet$	 & 26.8	 & 0.0	 & 0.781	 & 0.0  & u  \\  
NGC 6705	 & ASAS\,J184741-0654.4	 & $\bullet$ & $\bullet$ & $1.3\sigma$ &
$\bullet$ & $\circ$ & $\circ$	 & 34.6	 & 0.0	 & 0.778	 & 0.0  & i  \\
King 4	 & GO Cas	 & $\bullet$ & $\circ$ & $\bullet$ & $\bullet$ & $\circ$ & $1.3\sigma$	 & 21.7	 & 0.0	 & 0.684	 & 0.0  & i \\  
Berkeley 60	 & BF Cas	 & $\bullet$ & $\circ$ & $\bullet$ & $\bullet$ & $\circ$ & $1.1\sigma$	 & 33.9	 & 0.0	 & 0.672	 & 0.0  & i  \\  
Toepler 1	 & GI Cyg	 & $1.0\sigma$ & $\circ$ & $\circ$ & $\circ$ & $\circ$ & $\bullet$	 & 18.7	 & 0.0	 & 0.568	 & 0.0  & i \\  
Feinstein 1	 & U Car	 & $\bullet$ & $1.6\sigma$ & $\bullet$ & $\bullet$ & $\circ$ & $\bullet$	 & 20.8	 & 0.0	 & 0.52	 & 0.0  & i \\  
ASCC 61	 & VY Car	 & $\bullet$ & $\circ$ & $\bullet$ & $\bullet$ & $\circ$ & $2.0\sigma$	 & 21.9	 & 0.0	 & 0.385	 & 0.0  & i \\  
\hline
NGC 129	 & V379 Cas	 & $\circ$ & $\bullet$ & $\bullet$ & $\bullet$ & $\circ$ & $\bullet$	 & 20.3	 & 0.0	 & 0.896	 & 0.0  & y  \\  
Ruprecht 18	 & VZ CMa	 & $\circ$ & $\circ$ & $\bullet$ & $1.2\sigma$ & $\bullet$ & $\bullet$	 & 9.2	 & 0.0	 & 0.592	 & 0.0  & i \\  
Berkeley 82	 & ASAS\,J190929+1232.8	 & $\circ$ & $\circ$ & $\bullet$ & $\bullet$
& $\circ$ & $1.3\sigma$	 & 10.9	 & 0.0$^{*}$	 & 0.549	 & 0.0 & i  \\
Ruprecht 118	 & ASAS\,J162811-5111.9	 & $\circ$ & $\circ$ & $\bullet$ &
$1.1\sigma$ & $\circ$ & $\bullet$	 & 22.2	 & 0.0$^{*}$	 & 0.536	 & 0.0  & i \\
Hogg 12	 & GH Car	 & $\circ$ & $\circ$ & $\bullet$ & $1.1\sigma$ & $\circ$ & $1.1\sigma$	 & 9.8	 & 0.0	 & 0.503	 & 0.0  & i  \\  
NGC 6649	 & ASAS\,J183652-0907.1	 & $\circ$ & $\bullet$ & $\bullet$ &
$1.1\sigma$ & $\circ$ & $1.3\sigma$	 & 36.7	 & 0.0$^{*}$	 & 0.469	 & 0.0  & i  \\
Trumpler 9	 & ASAS\,J075503-2614.3	 & $\circ$ & $\circ$ & $1.1\sigma$ &
$1.1\sigma$ & $\circ$ & $\bullet$	 & 15.2	 & 0.0	 & 0.463	 & 0.0  & i  \\
NGC 2345	 & ASAS\,J070911-1217.2	 & $\circ$ & $\circ$ & $\bullet$ & $1.8\sigma$
& $\circ$ & $\bullet$	 & 36.6	 & 0.0	 & 0.317	 & 0.0  & i \\
\hline
Ruprecht 100	 & NSV 19202	 & $\circ$ & $\circ$ & $\circ$ & $\circ$ & $\circ$ & $\circ$	 & --	 & 0.837$^{*}$	 & 1.0	 & 0.837  & i \\  
FSR 1595	 & NSV 18905	 & $\circ$ & $\circ$ & $\circ$ & $\circ$ & $\circ$ & $\circ$	 & --	 & 0.821$^{*}$	 & 1.0	 & 0.821  & i \\  
Dolidze 53	 & V415 Vul	 & $\circ$ & $\circ$ & $\circ$ & $\circ$ & $\circ$ & $\circ$	 & --	 & 0.759$^{*}$	 & 1.0	 & 0.759  & i \\  
Ruprecht 100	 & TY Cru	 & $\circ$ & $\circ$ & $\circ$ & $\circ$ & $\circ$ & $\circ$	 & --	 & 0.671$^{*}$	 & 1.0	 & 0.671  & i \\  
SAI 116	 & NSV 18942	 & $\circ$ & $\circ$ & $\circ$ & $\circ$ & $\circ$ & $\circ$	 & 3.4	 & 0.621$^{*}$	 & 1.0	 & 0.621  & i \\  
Dolidze 34	 & TY Sct	 & $\circ$ & $\circ$ & $\circ$ & $\circ$ & $\circ$ & $\circ$	 & --	 & 0.547	 & 1.0	 & 0.547  & i \\  
\hline 
\end{tabular}
\label{tab:newCombos} 
\end{table*} 

\subsubsection{New candidate CCs}\label{sec:newCCs}

\paragraph{SX\,Car and ASCC\,61}
The $4.86$\,d Cepheid SX\,Car is seen to be co-moving with ASCC\,61 in proper
motion. Unfortunately, no average cluster RV is known. Parallax
and age, however, agree very well between cluster and Cepheid, lending support 
to the hypothesis of membership with $P(B|A) = 92\,\%$. An
in-depth analysis of the cluster, including its mean RV, is of the essence, 
since the cluster is located in a crowded field (K$05$). 
We tentatively consider SX\,Car as a member of ASCC\,61.

\paragraph{ASAS\,J182714-1507.1 and Kharchenko\,3}
The 5.5\,d fundamental-mode pulsator ASAS\,J182714-1507.1 = TYC\,6266-797-1 lies
at a large separation of $71$' from cluster center, which translates to
approximately $44$\,pc at the  cluster's estimated heliocentric distance of
$2.1$\,kpc and casts some doubt on possible membership.

Reddening for cluster and Cepheid (estimated from 2MASS photometry, see
Sec.\,\ref{sec:ebv}) are in excellent agreement, while parallax and age are
also consistent with membership. Proper motion does not exceed its error bars
and it thus of limited constraining power in this case. Detailed observational
follow-up is warranted for both Cepheid and cluster.

\paragraph{S\,Mus and ASCC\,69} The membership constraints considered (all but
[Fe/H]) for
the binary Cepheid S\,Mus \citep[e.g.][]{2005MNRAS.362.1167P} and ASCC\,69 yield
a very high likelihood of $88\,\%$. 
ASCC\,69 is a sparsely populated cluster for which K$05$ list merely 
twelve 1-$\sigma$ members. Cluster radius and center coordinates may
therefore be rather imprecise. Furthermore, radial velocities are only of
limited value as membership constraints, since the average cluster RV is based
on only 2 stars. However, proper motion clearly indicates that cluster and
Cepheid are co-moving. We tentatively accept S\,Mus as a cluster member and
stress the need for a detailed study of its candidate host cluster ASCC\,69.

\paragraph{UW\,Car and Collinder\,220} 
This new Combo yields a likelihood of $84\,\%$ from all membership
constraints but [Fe/H]. The parallax of the Cepheid is estimated by the PLR, cf.
Sec.\,\ref{sec:Clplx}. The most compelling evidence of membership comes from
proper motion, while the large on-sky separation translates into a distance of
$47$\,pc from cluster center assuming the cluster's distance. A detailed review
of the cluster's parameters, as well as a better parallax estimate of the
Cepheid would help to conclude on this Combo.

\paragraph{V379\,Cas and NGC\,129}
This high likelihood pair at large separation ($44'$ or 20\,pc at the
estimated distance to NGC\,129) has nearly vanishing proper motion, while the both RVs are
in excellent agreement and the ages are consistent. We obtain a likelihood of
$91\,\%$.
No parallax is computed, since V379\,Cas is an overtone pulsator.
Since NGC\,129 has another known member, DL\,Cas, 
we can compare the pulsational ages of the two Cepheids
and find both to be consistent within the uncertainties ($7.70\pm0.08$ for DL\,Cas
and $7.83\pm0.07$ for V379\,Cas). Furthermore, the iron
abundances of both Cepheids are close, as are their RVs (to within less than
1\,km\,s$^{-1}$). We further note that V379\,Cas and DL\,Cas have similar
reddening values, though $E(B-V)$ of V379\,Cas is slightly ($0.1$\,mag) higher
\citep[both]{2008MNRAS.389.1336K}.
We therefore tentatively consider V379\,Cas a member of NGC\,129's Halo, pending
a better distance estimate and additional membership constraints.  NGC\,129 is
thus particularly interesting, containing both a fundamental-mode and an
overtone pulsator, just as NGC\,6067.

\subsubsection{Unconstrained High-prior Combos}
In Tab.\,\ref{tab:newCombos}, we highlight six Combos with  
$P(A) > 50\%$ that have thus far not been studied for membership, and for which
no membership constraints were available. Hence, no likelihoods 
could be computed for these cases. We therefore suggest the following Combos for
observational follow-up:
V415\,Vul -- Dolidze\,53; TY\,Cru \& NSV\,19202 -- Ruprecht\,100;
NSV\,18942 -- SAI\,116; TY\,Sct -- Dolidze\,34; NSV\,18905 -- FSR\,1595.

\subsection{The Galactic Cepheid PLR revisited}\label{sec:GalPLR}
\begin{figure*}
\centering
\begin{tabular}{lr}
\includegraphics{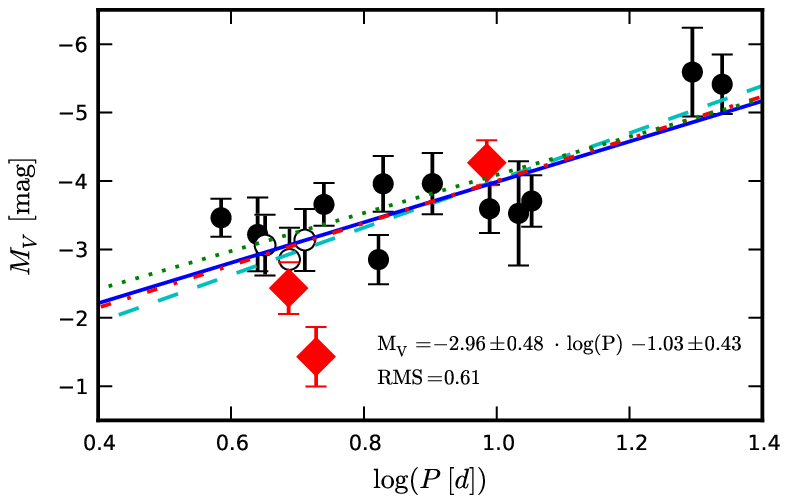}
\includegraphics{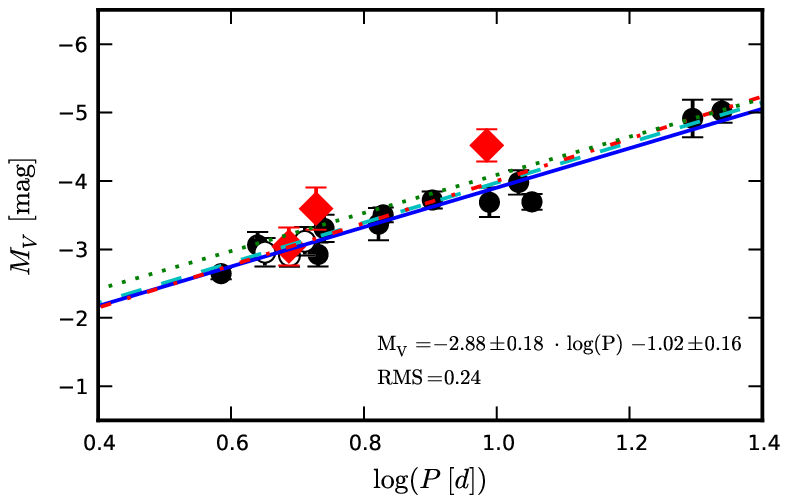}
\end{tabular}
\caption{Cepheid PLRs fitted to the `standard' (left panel) and `optimal'
(right panel) data sets. Solid circles indicate previously known CCs, open
circles highlight the three Cepheids in NGC\,7790. Solid red large diamonds
indicate the new \textit{bona-fide} fundamental-mode CCs S\,Mus,
SX\,Car, and UW\,Car included in the fit.
The dotted line shows the PLR by \citealt{2007AJ....133.1810B}, and the
dash-dotted line represents \citet{2003A&A...404..423T}. Solid and
dashed lines indicate weighted and non-weighted least-squares fits. An
accurate PLR calibration critically depends on accurate distance estimates from
detailed studies that include line-of-sight-variations of extinction.}        
\label{fig:plr}
\end{figure*}  

\begin{table*}
\centering
\begin{tabular}{llrrrrrr}
\hline
Cepheid & Cluster & $\langle m_V \rangle$ & $E(B-V)$ & $(V_0-M_V)$ & $R_{V,{\rm Cl}}$ & $\log{P}$ & $M_V$ \\
\hline
SU Cyg  & Turner 9 & 6.89 & $ 0.07 \pm 0.02 $ & $ 9.33 \pm 0.05 $ & $ 2.94 \pm 0.38 $ & 0.585 & $ -2.65 \pm 0.08 $ \\
CG Cas & Berkeley 58 & 11.37 & $ 0.69 \pm 0.01 $ & $ 12.40 \pm 0.12 $ & $ 2.95
\pm 0.20 $ & 0.640 & $ -3.07 \pm 0.19 $ \\
CE Cas B & NGC 7790 & 11.09 & $ 0.48 \pm 0.05 $ & $ 12.46 \pm 0.01 $ & $ 3.31 \pm 0.26 $ & 0.651 & $ -2.98 \pm 0.21 $ \\
SX Car$^{*}$ & ASCC 61 & 9.12 & $ 0.33 \pm 0.03 $ & $ 11.14 \pm 0.20 $ & $ 3.10 \pm 0.51 $ & 0.687 & $ -3.04 \pm 0.28 $ \\
CF Cas  & NGC 7790 & 11.15 & $ 0.48 \pm 0.03 $ & $ 12.46 \pm 0.01 $ & $ 3.33 \pm 0.26 $ & 0.688 & $ -2.91 \pm 0.16 $ \\
CE Cas A & NGC 7790 & 10.94 & $ 0.48 \pm 0.05 $ & $ 12.46 \pm 0.01 $ & $ 3.33 \pm 0.26 $ & 0.711 & $ -3.12 \pm 0.21 $ \\
UW Car$^{*}$ & Collinder 220 & 9.46 & $ 0.46 \pm 0.01 $ & $ 11.63 \pm 0.20 $ & $ 3.10
\pm 0.51 $ & 0.728 & $ -3.60 \pm 0.31 $ \\
CV Mon & vdBergh 1 & 10.33 & $ 0.68 \pm 0.05 $ & $ 11.08 \pm 0.07 $ & $ 3.20 \pm
0.04 $ & 0.731 & $ -2.93 \pm 0.18 $ \\
V Cen  & NGC 5662 & 6.87 & $ 0.25 \pm 0.05 $ & $ 9.31 \pm 0.02 $ & $ 3.47 \pm
0.38 $ & 0.740 & $ -3.32 \pm 0.20 $ \\
BB Sgr & Collinder 394 & 6.91 & $ 0.29 \pm 0.05 $ & $ 9.38 \pm 0.10 $ & $ 3.10
\pm 0.51 $ & 0.822 & $ -3.37 \pm 0.24 $ \\
U Sgr  & IC 4725 & 6.72 & $ 0.39 \pm 0.02 $ & $ 8.93 \pm 0.02 $ & $ 3.32 \pm 0.21 $ & 0.829 & $ -3.52 \pm 0.11 $ \\
DL Cas & NGC 129 & 8.98 & $ 0.46 \pm 0.02 $ & $ 11.11 \pm 0.02 $ & $ 3.46 \pm 0.22 $ & 0.903 & $ -3.73 \pm 0.13 $ \\
S Mus$^{*}$ & ASCC 69 & 6.13 & $ 0.21 \pm 0.02 $ & $ 10.0 \pm 0.20 $ & $ 3.10 \pm 0.51 $ & 0.985 & $ -4.52 \pm 0.24 $ \\
S Nor  & NGC 6087 & 6.41 & $ 0.12 \pm 0.05 $ & $ 9.65 \pm 0.03 $ & $ 3.74 \pm
0.85 $ & 0.989 & $ -3.70 \pm 0.22 $ \\
TW Nor & Lyng\aa\ 6 & 11.66 & $ 1.24 \pm 0.03 $ & $ 11.51 \pm 0.08 $ & $ 3.33
\pm 0.10 $ & 1.033 & $ -3.96 \pm 0.18 $ \\
V340 Nor & NGC 6067 & 8.41 & $ 0.32 \pm 0.02 $ & $ 11.03 \pm 0.01 $ & $ 3.36 \pm
0.29 $ & 1.053 & $ -3.68 \pm 0.11 $ \\
RU Sct & Trumpler 35 & 9.53 & $ 0.92 \pm 0.03 $ & $ 11.58 \pm 0.18 $ & $ 3.10
\pm 0.20 $ & 1.295 & $ -4.91 \pm 0.27 $ \\ 
WZ Sgr & Turner 2 & 8.09 & $ 0.62 \pm 0.02 $ & $ 11.26 \pm 0.10$ & $ 3.00 \pm
0.20 $ & 1.339 & $ -5.02 \pm 0.17 $ \\
\hline
\end{tabular}
\caption{Parameters adopted for the `optimal' set used in Eq.\,\ref{eq:absmag} and the right panel of Fig.\,\ref{fig:plr}. Newly-identified \textit{bona fide} Combos employed in the fit are marked with an asterisk next to the Cepheids identifier. True distance moduli of clusters, $\left(V_0 - M_V\right)$, and absorption-relevant parameters were adopted according to the criteria specified in the text. We adopt $0.04$\,mag as the uncertainty on $\langle m_V \rangle$.}
\label{tab:cldistmod}
\end{table*}

Let us now employ our \textit{bona-fide} CC sample to revisit the calibration
of the Galactic Cepheid period-luminosity relationship. It represents an ideal sample to this
end, due to the high confidence we can have in cluster membership,
though small statistics and a lack of long-period calibrators
will limit the precision attainable.

A Cepheid's absolute V-band magnitude, $M_V$, is determined using the true distance modulus of the cluster, $\left( V_0 - M_V \right)_{\rm Cl}$, the Cepheid's mean magnitude, $\langle m_{V} \rangle$, 
the ratio of total-to-selective extinction towards the cluster, $R_{V,{\rm
Cl}}$, and the Cepheid's color excess, $E(B-V)$, as
\begin{equation}
M_{V} = \langle m_{V} \rangle - \left( V_0 - M_V \right)_{\rm Cl} - R_{V,{\rm
Cl}} E(B-V) \, ,
\label{eq:absmag}
\end{equation}
where quantities refer to the Cepheid, unless subscripted by `Cl'.

As described in Sec.\,\ref{sec:LitData}, we compile cluster data from
D02 for our membership analysis. These data, however, do not take into account
line-of-sight dependencies of reddening, or non-canonical values of $R_V$. In
the following, we refer to the set of data compiled for our membership analysis
as the `standard' set.

In an attempt to improve accuracy, we compile more accurate data from
detailed studies of the host clusters. True distance moduli based on
ZAMS-fitting were taken from  T$10$ and \cite{2007ApJ...671.1640A}, giving
preference to the estimates with the smallest uncertainties.
For the two ASCC clusters, we use the de-reddened values by
\cite{2005A&A...440..403K}, and for Collinder\,220 we rely on the data in D02. 
For reddening, \cite{2007ApJ...671.1640A} provide a convenient way to calculate
$R_V$ for some lines of sight in our sample as $R_V = R_{V,0} + 0.22(B-V)_0$,
with $R_{V,0}$ tabulated for the cluster and taking into account the intrinsic
color of the Cepheids.
For the two ASCC clusters and Collinder\,220, we employ the canonical $R_V =
3.1 \pm 0.2$ since no other estimate is available. For Turner\,9, Berkeley\,58,
and van den Bergh\,1, we use the distance moduli and $R_V$ values from
\cite{1997AJ....113.2104T,2008MNRAS.388..444T,1998AJ....115.1958T},
respectively. For the long period Cepheid hosts, Trumpler\,35 and Turner\,2, we
employ the data published in \cite{1980ApJ...240..137T} and
\cite{1993ApJS...85..119T}. Cepheid mean V-magnitudes and $E(B-V)$ were
compiled as described in Sec.\,\ref{sec:LitData}. Note that ASAS\,J$182714-1507.1$ was not included in this calibration, since the data compiled were not of sufficient quality.
The values thus compiled are listed in Tab.\,\ref{tab:cldistmod}, and
we refer to this data set as the `optimal' one. 

Figure\,\ref{fig:plr} shows the fits to both the `standard' (left panel) and the
`optimal' (right panel) data sets. In both figures, four straight lines indicate
the PLR calibrations by \citet[red dash-dotted]{2003A&A...404..423T},
\citet[green dotted, lower zero-point]{2010OAP....23..119T}, as well as our
non-weighted (dashed cyan), and weighted (solid blue) least-squares fits to the
data. The large scatter (RMS=0.61) in the `standard' data set is striking.
Contrastingly, the `optimal' set is much better indeed, with an RMS
of 0.24\,mag. Hence, it is evident that the cluster data compiled for the membership analysis is rather limited in precision and sometimes also accuracy, a fact we already encountered when computing membership probabilities, see for instance the case of WZ Sgr in Sec.\,\ref{sec:WZSgr}. Thus, there remains a need for detailed and deep photometric studies of open cluster parameters, and in particular reddening.

We determine the uncertainties on the `optimal' set by linear
regression and obtain:  
\begin{equation}  \langle M_V \rangle = -\left(2.88 \pm 0.18\right) \log{P} -
\left(1.02 \pm 0.16\right) \,.
\label{eq:myplr}
\end{equation}
Despite the reasonable formal uncertainties of our `optimal' fit, our
solution should not be considered definitive. 
The result of the fit is very sensitive to the absolute magnitude estimates of
the extreme points at short and long periods, and the $M_V$ estimates of the new 
candidates are
clearly too crude at this point. Furthermore, despite our preference for cluster
literature with the smallest uncertainties, there is no guarantee that the most
accurate cluster data was employed; there appears to exist too little consensus
on some clusters in the literature, e.g. for Lyng\aa\ 6, $\left( V_0 - M_V
\right)_{\rm Cl}$ differs by 0.39\,mag between T10 and
\cite{2007ApJ...671.1640A}, exceeding the published combined uncertainties by a
factor greater than 3. Binarity of the Cepheids was not accounted
for, since the fit is dominated mainly by the cluster parameters. Yet, we note
that our result is consistent with the calibration by T10 ($\langle M_V \rangle=
-\left(2.78 \pm 0.12\right) \log{P} - \left(1.29 \pm 0.10\right)$), which is to
be expected due to the significant overlap in the sample of CCs used for this
calibration.

\section{Discussion}\label{sec:Discussion}
\subsection{Constraining Power and Limitations of Membership Constraints}
\subsubsection{The Prior P(A)}
The form of the prior was motivated by 
radial density profiles of star clusters, see Sec.\,\ref{sec:radii}.
However, the degree with which the distribution of stars in the cluster is known varies greatly between clusters, and deviations from circular cluster shapes were ignored for internal consistency. Furthermore, crowding, great distances ($\sim$\,kpc) to host cluster candidates, differential reddening, sparsity, etc. all conspire to complicate the definition of cluster radii. It is thus not surprising that the prior does not perform extremely well as a membership constraint if taken at face value. Nevertheless, it does help to separate the interesting cases from the majority of null matches, since it reduces the question of proximity on the projected sky to a single number that contains information on the density of the cluster, since both cluster radii are used in its definition.

\paragraph{Chance Alignment}
Among our (\textit{bona fide} or inconclusive) CCs, $7 - 9$ are found to lie
within $r_c$, and $8 - 17$ at $r_c < R < r_{\rm lim}$. Within the original
cross-match, $25$ Combos are matched
within $r_c$ and $520$ within $r_c < R < r_{\rm lim}$. Thus we can estimate the
rate of chance alignment within the core radius to be approximately $3:1$.
At separations inferior to $r_{\rm lim}$, this ratio increases to between $20:1$ and $35:1$, depending on whether inconclusive cases are counted, or not.

Note that $8$ \textit{bona-fide} CCs lie outside $r_{\rm lim}$, $6$
of which are located within two $r_{\rm lim}$; EV\,Scuti's separation from NGC\,6664's center is $2.6\,r_{\rm lim}$, and that of V379\,Cas from NGC\,129 is $2.7\,r_{\rm lim}$.

\subsubsection{The Likelihood P(B$\mathbf{\vert}$A)}\label{sec:disc:likelihood}
\begin{figure}
\includegraphics{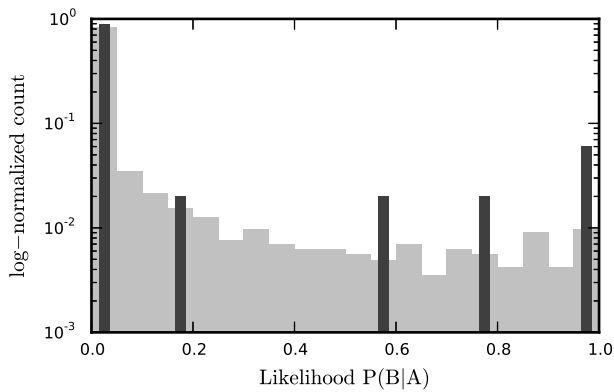}
\caption{Log-normalized histogram of likelihoods. Light gray, broad
bars:
Combinations with 3 or more parameters used for P(B$|$A). Dark gray, slim bars:
Combinations with all membership constraints. The more membership constraints
are employed, the more separated are high and low likelihoods, i.e. the better
constraint is  membership.}  
\label{fig:constrainthist}
\end{figure}

Intuitively, the greatest set of constraints used provides the tightest constraints on membership for any cluster-Cepheid combination (Combo).
Figure\,\ref{fig:constrainthist} illustrates this. It shows
a logarithmic normalized histogram of likelihoods for two cases: more than 3
constraints used to calculate P(B$|$A) (light gray distribution); all
constraints used to calculate P(B$|$A) (dark gray slim bars). However,
the constraining power of a given set of constraints is not merely a function of its size. Here we discuss the membership constraining power of the different constraints used to calculate likelihoods.

\begin{figure}
\includegraphics{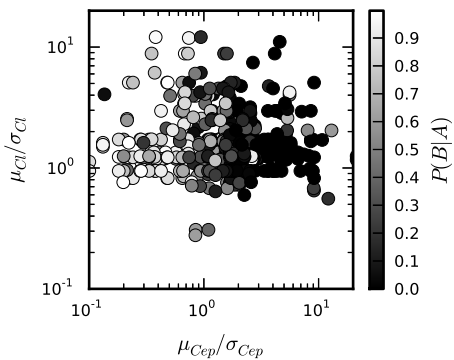}
\caption{Likelihoods computed as a function of proper motion. Abscissa: proper
motion of Cepheid divided by the squared-summed uncertainties, i.e. $\mu =
\sqrt{(\mu_\alpha^*)^2 + \mu^2_\delta}$; $\sigma =
\sqrt{\sigma\left(\mu_{\alpha}^{*}\right)^2 + \sigma\left(\mu_\delta\right)^2}$.
Ordinate: same for Cluster.}    
\label{fig:disc_pm}
\end{figure}

\textbf{Proper motions} can be very effective at ruling out
membership if the motion clearly exceeds the uncertainties on the measurement. 
This is the case for a cluster cross-matched with a background Cepheid, for instance. 
However, for a significant fraction of Combos, the proper motion vector's magnitude was smaller than the uncertainty of the measurement, thus effectively not constraining membership, see Fig\,\ref{fig:disc_pm}.
If the magnitude surpasses the uncertainties by at least a factor of 3, proper motion serves as a reliable constraint. For the majority of Combos that fulfill this criterion, membership tends to be excluded.

\textbf{Distance} is a potentially very strong membership constraint, since 
intuitively, a Cepheid that occupies the same space volume as a cluster should be a member.
However, cluster distances can be subject to large systematic uncertainties due to parameter degeneracy (distance, age, reddening), model-dependence (rotation, etc.), or previous distance estimates to, e.g., the Pleiades. 
Furthermore, implicit assumptions on cluster membership
can significantly impact the distance determined, 
especially for relatively young clusters that harbor few stars around the Main-sequence turn-off and few or no red giants. 
Very detailed studies of open clusters, and in particular of the line-of-sight extinction, are required for improvement in this domain, as is shown in Sec.\,\ref{sec:GalPLR}, or demonstrated by the discussion of CV\,Mon's membership in van den Bergh\,1 (cf. Sec.\,\ref{sec:cvmon}).

\textbf{Radial velocities} have the potential to provide very tight membership
constraints, since the RV dispersion within open clusters can be significantly below 1\,km\,s$^{-1}$ \citep{2007A&A...472..657L}, approaching the measurement precision on $v_\gamma$ for non-binary Cepheids. 
However, estimates of average cluster RVs are usually based on only a few stars, see Fig.\,\ref{fig:disc_rv}. 
In fact, approximately half of all clusters with `known' average RVs are based on measurements of two stars or less, and strong selection effects (e.g. toward late-type stars) can severely impact the estimate.
Since Cepheids are very bright and of late spectral type, it is rather likely that a cluster RV is in part based on measurements that include 
a Cepheid. 
We therefore highlight the need to observe more radial velocities of
upper main sequence stars in clusters in order to ensure the most accurate
estimates of average cluster RV. Alternatively, larger telescopes may
observe the much fainter lower main sequence. The Gaia-ESO public spectroscopic
survey \cite{2012Msngr.147...25G} will soon provide precise RVs for a
large number of clusters and thus can improve the reliability of a future study
similar to the present one. 

The \textbf{iron abundances} compiled here were, arguably, of limited use as
membership constraints, since data inhomogeneity, the limited number of cluster stars
used for determining the cluster average, and differences in the solar
reference values used are at the same order of magnitude as the range of iron
abundances found in the sample considered (that lies within the young metal-rich
Galactic disk). For a few cases, however, the iron abundance did further
strengthen the interpretation of excluded membership.

\textbf{Age} as a membership constraint quantifies valid evolutionary considerations that are established empirically. It is a particularly useful membership constraint, since it is readily available for clusters as well as for Cepheids (from period-age relations), especially when few other constraints are available. 
However, ages for both kinds of objects are subject to model-dependence, and the accuracy of the values inferred is difficult to quantify.

All of the above quantities have their own peculiarities, and thus no single one
can be named the `best' membership constraint. Instead, the greatest constraining power resides in the combination of all the data, as is seen in the
\textit{bona-fide} CCs identified in this work, as well as in
Fig.\,\ref{fig:constrainthist}.

\subsubsection{Membership Probabilities P(A$\mathbf{\vert}$B)}
\begin{figure}
\includegraphics{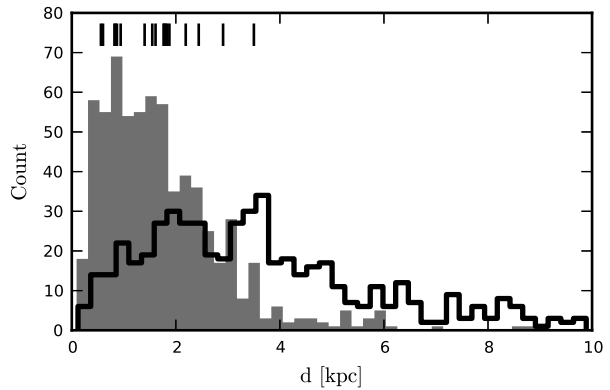}
\caption{Distance distribution of known open
clusters in log-age range [7.0, 8.5] (gray filled histogram, high
peak at small d) and of Cepheids at low Galactic latitudes, $|b| \leq
10^{\circ}$ (black step histogram). Distances of \textit{bona-fide} CCs from
this work indicated by short black lines at the top.}   
\label{fig:OCCepHist}
\end{figure}

We computed $P(A\vert B)$ simply as the product of the prior and the likelihood,
leaving out the normalization term P(B) that would in principle be required, see
Eq.\,\ref{eq:bayes}. However, in order to make full use of the  $P(A\vert B)$
values computed, P(B) should not be neglected. Given that the incompleteness for
open clusters is quite significant at heliocentric distances greater than, say,
1\,kpc, we did not currently see this as feasible, cf.
Fig.\,\ref{fig:OCCepHist}.

\subsection{Incompleteness}
Despite our aim to maximize the number of clusters and Cepheids
considered, only 23 \textit{bona fide} CCs were identified. This is a
result of the apparent difference in sample completeness for clusters and Cepheids.
Fig.\,\ref{fig:OCCepHist} shows the histograms of heliocentric open cluster
distances for clusters withing the appropriate age range $\left(
\log_{10}{\left( age [yr]\right)} = [7.0, 8.5] \right)$, taken from D02
as the grey filled distribution (peak at smaller d), and the Cepheid
distances compiled (cf. Sec.\,\ref{sec:CepheidDistances}, black step histogram).
It is evident that the detection rate of open clusters stalls at
distances greater than 1\,kpc, probably since their identification against the
field becomes increasingly difficult. Cepheids, on the other hand, are
detectable at much greater distances, thanks to their high luminosity and
characteristic brightness variations.

However, a cluster's probability of hosting a Cepheid is governed by stellar
evolution and star formation. For a given distribution of stellar masses, only a
small fraction will become seen as Cepheids during their lifetime (stars with
masses between $\sim\, 4-11 M_\odot$). Due to the nature of the IMF, these
intermediate mass stars constitute a small fraction of the total number of
cluster stars, and few such intermediate-mass stars are present in typical
(small) open clusters. Furthermore, only a fraction (perhaps 10-20\,\%) of a
suitable star's lifetime during the core Helium burning phase is spent on the
blue loops, and even less within the instability strip.     
As a result, few CCs are known. The inverse problem of finding host clusters
around known Cepheids \citep[e.g.]{1993ApJS...85..119T,1998AJ....116..274T,2012ApJ...748L...9M,2012MNRAS.421.1040M}, however, can be interesting, e.g. to constrain the
survival rate of open clusters. Such studies can be very successful, see
e.g. the ``missing Combos'' in Sec.\,\ref{sec:missedCCs} and the Cepheids
belonging to OB associations listed in T10. 

\subsection{Implications for the Distance Scale}
We are aware of the fact that our calibration, although performed
on an `optimal' set, remains inhomogeneous in the true distance moduli
employed and the treatment of extinction. It is furthermore based on
V-band data obtained in multiple different passbands with varying post
processing techniques. These are the main limitations of the data employed.
Furthermore, the number of CCs employed in the fit (18) is not
very large, and the distance to NGC\,7790 enters the fit three
times, since its three CCs are included in the fit. Finally, 
the fit obtained is very sensitive to outliers, due to the lack of long-period
calibrators.

We advocate that the `optimal' sample presented, although incomplete (missing
host clusters), forms an ideal sub-set for PLR
calibrations, since all cluster memberships were self-consistently evaluated.
However, to improve upon the calibration of the cluster-based PLR, two
things would be particularly useful: a detailed homogeneous deep photometric study of
the host clusters that includes careful treatment of extinction,
and observational follow-up of the inconclusive CC candidates identified.
Given the large discrepancies that are found between recent Galactic PLR
calibrations (and their zero-points), such an observational campaign would be
very desirable. 


\section{Conclusion}\label{sec:conclusion}
Focusing on Cepheids, we have performed an all-sky cluster membership census.
Our analysis considers an up to 8-dimensional membership space that includes
spatial and kinematic information, as well as parent population parameters (age,
iron abundance). Although in some ways limited by data inhomogeneity and
incompleteness, we identify 23 
\textit{bona fide} cluster Cepheids, including most canonical CCs accessible
within our sample, as well as 5 new ones, and multiple additional CC
candidates of interest.

The newly identified CCs are: SX\,Car in ASCC\,61, S\,Mus in ASCC\,69,
UW\,Car in Collinder\,220, ASAS\,J182714-1507.1 in Kharchenko\,3 (fundamental-mode pulsators), and V379\,Cas in NGC\,129 (overtone pulsator).
The cluster membership of these candidates must have escaped previous discovery,
since most of the host clusters are not very well-studied, and the Cepheids are
located outside the cluster cores, cf. Sec.\,\ref{sec:newCombos}.

Since we can rank candidates according to membership probabilities, we consider
our \textit{bona-fide} CC sample ideal for calibrating the Galactic Cepheid
PLR. However, data inhomogeneity, large uncertainties on cluster
parameters, and a lack of long-period calibrators unfortunately limit the
precision of the calibration we perform.
We therefore highlight the need for observational
campaigns dedicated to the host clusters of our \textit{bona-fide} CC sample,
as well as other interesting CC candidates identified.

The limitations that our work suffers due to inhomogeneous and incomplete data will be significantly reduced in the near future, thanks to the Gaia space mission. Specifically, Gaia will improve our study in the following ways:
\begin{itemize}
\item Thousands of new Cepheids \citep{2000ASPC..203...71E,2011A&A...530A..76W}
will be discovered. 
\item Accurate absolute trigonometric parallaxes of Cepheids will become
available up to distances of 6-12\,kpc, depending on extinction
\citep{2012Ap&SS.341..207E}, thereby enabling a direct calibration of the
Galactic period-luminosity-relationship similar to the one performed in the
seminal paper by \cite{1997MNRAS.286L...1F}. This will remove our partial
dependence on existing PLR calibrations when determining cluster membership.
Accurate parallaxes to longer period Cepheids will also be obtained, thereby
significantly improving the distance estimates to extragalactic Cepheids.      
\item The accuracy of proper motions will be improved by orders of magnitude,
moving from mas\,yr$^{-1}$ to tens of $\mu$as\,yr$^{-1}$, see \cite{2010IAUS..261..296L} and the Gaia Science Performance website\footnote{\texttt{http://www.rssd.esa.int/index.php?project=GAIA\& page=Science\_Performance\#chapter1}}.
\item Homogeneous radial velocities (via the RVS instrument) and metallicity estimates (via the spectrophotometric instruments) will be available as membership constraints for a great number of Cepheids.
\item Homogeneous metallicity estimates can be obtained through spectrophotometry-photometry \citep{2012MNRAS.426.2463L}.
\item Thousands of new open clusters \citep{ESA2000} will be discovered, forming a more or less complete census of open clusters to distances up to 5\,kpc. As was shown in Fig.\,\ref{fig:OCCepHist}, the distribution of Cepheids is still increasing at these distances. Hence, one may expect to find many more CCs in these parts of the Galaxy.
\item Down to magnitude 20, all-sky homogeneous multi-epoch photometry and
colors will be obtained that will include all the \textit{bona-fide} cluster
Cepheids mentioned in this work.
\item Known clusters will be mapped in unprecedented detail, and intra-cluster dynamics will be accessible to determine membership.
\end{itemize}
Gaia's data homogeneity will significantly improve error budgets, since no offsets in instrumental zero-points (e.g. in RV) will have to be taken into account. The constraining power in terms of membership will thus be augmented considerably. Correlations between parameters, e.g. proper motion, parallax, and RV, can be determined self-consistently  and accounted for \citep[cf.][]{2007ASSL..350.....V}. Such factors will make the Gaia era particularly exciting for work such as this.

\section*{Acknowledgments} Many heartfelt thanks to the observers who
contributed to the Cepheid radial velocity campaigns, in particular
to:
Lovro Palaversa, Mihaly V\'aradi, and Pierre Dubath. We gratefully acknowledge
useful discussions with and comments received from Maria S\"uveges, Berry
Holl, David G. Turner, Daniel Majaess, Laszlo Szabados, Imants Platais, Michael
W. Feast, and the referee, C. David Laney; all of these helped to improve the
manuscript.\\
This research has made use of:
NASA's Astrophysics Data System Bibliographic Services; the SIMBAD database and
VizieR catalogue access tool (cf. A\&AS 143, 23), operated at CDS, Strasbourg,
France; the International Variable Star Index (VSX) database, operated at AAVSO,
Cambridge, Massachusetts, USA; the VO-tool
TOPCAT\footnote{\texttt{http://www.star.bris.ac.uk/$\sim$mbt/topcat/}}, see
\cite{2005ASPC..347...29T}; the WEBDA database, operated at the Institute for
Astronomy of the University of Vienna; other online databases that provide
Cepheid data, see article body.

\bibliographystyle{mn2e} 
\bibliography{mn-jour,cepcat}    


\appendix\label{appendix}
\section{Details on Individual Cluster Cepheid Combinations}
\subsection{Literature Combos}\label{sec:litAppendix}
\subsubsection{Inconclusive Combos}

\paragraph{CS\,Vel and Ruprecht\,79} 
Membership of CS\,Vel in Ruprecht\,79 was thoroughly discussed by
\cite{1976ApJ...209..130H} who 
credited \cite{1966ATsir.367....1T} with first suggesting this particular
combination. It has since been studied multiple times, e.g. by
\cite{1987MNRAS.229...31W} and T$10$. Due to the sparse nature of the cluster,
its reality as such was doubted by \cite{2005ApJS..161..118M}, who conclude that
Ruprecht\,79 rather be a hole in the dust of the Sagittarius-Carina spiral arm
than a physical open cluster.

None of the data from D02 are fully consistent with cluster membership
for CS\,Vel, see Tab.\,\ref{tab:literature}, and we calculate a likelihood of
not even $1\,\%$, which contrasts the Cepheid's location inside
$r_{\rm c}$.
The Cepheid's color excess from \cite{2007MNRAS.377..147L} agrees with the
cluster value, however, and RV is not very far off. The cluster data from D02
and the uncertain existence of the cluster would suggest unlikely membership. 
However, the difficult parameter determination for this sparse cluster
(candidate) means that only a very detailed study of this region can reliably
conclude on membership.

\paragraph{V1726\,Cyg and Platais\,1} V1726\,Cyg's membership results for
Platais\,$1$ are based on separation, proper motion, radial velocity, and age.
The star was first considered for cluster membership by
\cite{1979ATsir1049....4P} and \cite{1994AJ....107.1796T}, and is still
often considered a \textit{bona fide} member of Platais\,1
\citep{2001PASP..113..715T,2006Obs...126..207T}.
However, \cite{1986NInfo..61...89P} concludes that the existence of this cluster
is uncertain.

Of the membership constraints compiled, only proper motion
does not significantly differ between cluster and Cepheid,
although the magnitude of the Cepheid's motion is a bit larger than that
of the cluster. Radial velocity differs significantly, with the cluster receding
more than $10$\,km\ s$^{-1}$ faster than the Cepheid
\citep{2008AJ....136..118F}. In addition, the cluster's age is significantly
higher than the Cepheid's.
We do not calculate a PLR-based parallax, since V$1726$\,Cyg may or may
not be an overtone pulsator; the Fernie database places V1726\,Cyg at
approximately $2$\,kpc, while D02 list 1.3\,kpc for Platais\,$1$.
 
Unfortunately, there are great differences in cluster parameters to be
found in the literature. For instance, K$12$ list $3.5$\,kpc as the
cluster's distance, which would exclude membership and be consistent with the
higher proper motion of a foreground Cepheid. Interestingly, however, K12's 
average cluster RV ($-15.4$\,km\ s$^{-1}$) would agree with the Cepheid's,
although it is based on 21 stars in a poorly populated and supposedly
distant (3.5\,kpc in K12) cluster, rendering this estimate suspicious. 

In summary, the constraints compiled here would tend to indicate
non-membership.
However, the significantly discrepant cluster distances and  radial velocities 
from different references leave considerable doubt as to whether the
cluster values adopted here are accurate. In light these issues and
the possible non-existence of the cluster, we cannot conclude on membership of
this Combo.

\subsubsection{Unlikely Literature Combos}

\paragraph{GH\,Car in Trumpler\,18 or Hogg\,12} Membership of the
spectroscopic binary overtone pulsator GH\,Car
\citep{2013MNRAS.430.2018S} in cluster Trumpler\,18 was first proposed by  \cite{1990A&AS...86..209V} and then
called into question by \cite{2000A&AS..146..251B} who recommended radial
velocity follow-up to draw a firmer conclusion. We compute a low likelihood of
$14\,\%$ based on proper motion, age, and RV.
No parallax was calculated, since GH\,Car is an overtone pulsator.
However, the Cepheid's distance listed in the Fernie database ($2.2$\,kpc) is
significantly larger than Trumpler\,18's ($1.4$\,kpc). Furthermore, the Cepheid
is reddened by $0.1$\,mag more than the cluster, which is consistent with a
greater distance to the Cepheid. The average cluster RV is based on the
measurements of a single star, and does not agree with that of the Cepheid. In
short, membership of GH\,Car in Trumpler\,18 is unlikely.

However, our analysis identifies Hogg\,12 as an alternative host
cluster for GH\,Car. Based on proper
motion and age, we compute a likelihood of $50\,\%$ for this Combo. Proper motion is a very good match, and clearly
detected. Finally, reddening for the cluster and the Cepheid are nearly
identical, and the Cepheid's literature distance from the Fernie database
matches the cluster's distance very well. Follow-up is required to confirm this
option.

\paragraph{SZ\,Tau and NGC\,1647}\label{sec:sztau} The membership of SZ\,Tau in
the halo of NGC\,1647 was first considered by \cite{1964PZ.....15..242E} and later
studied in more detail by \cite{1992AJ....104.1865T} who concluded that
SZ\,Tau is a `coronal' member, based on star counts, reddening, radial velocity,
proper motion from \cite{1989AJ.....98..888F}, and assuming overtone pulsation
for SZ\,Tau. 

From parallax \citep{2011A&A...534A..94S}, radial velocity, proper
motion, and age, we compute a likelihood of membership of $5\,\%$, the main
discrepant constraints being radial velocity and proper motion from Tycho2
and Hipparcos \citep{2001A&A...376..441D,2007ASSL..350.....V}.
The vanishing prior could, of course, be consistent with coronal membership.
However, cluster membership based on proper motion was excluded by 
\cite{1996A&AS..118..277G} using 2220 stars measured on photographic plates, and
by \cite{2000A&AS..146..251B} using Hipparcos proper motions. We therefore
consider SZ\,Tau an unlikely member of NGC\,1647, although an ejection cannot be
excluded (Turner 2013, priv. comm.).

\subsubsection{Non-member Combos discussed in the Literature}
\paragraph{V442\,Car and NGC\,3496} V442\,Car was previously considered for
membership in NGC\,$3496$ by \cite{1995MNRAS.277..250B} who concluded it to be a
background star, based on  age and reddening. From proper motion and separation,
we come to the same conclusion. Furthermore, a rough distance estimate for a
$14^{\rm{th}}$ magnitude 5.5\,d Cepheid excludes membership in a cluster
located approx. 1\,kpc from the Sun.

\paragraph{UY\,Per and Czernik\,8 or King\,4}\label{sec:uyper}
\cite{1977PASP...89..277T} suggested that UY\,Per could be a member of
Czernik\,8 or King\,4. \cite{2010OAP....23..119T} again mentions the latter
combination. Our results, however, are inconsistent with membership in either
cluster, based on the constraints parallax, proper motion, and age. 

The `likelier' of the two Combos is King\,4, 
for which parallax and age are in relatively good agreement; the
Cepheid is slightly farther away and has larger reddening. Kinematically,
however, the cluster's vanishing proper motion is inconsistent
with membership of the rather fast moving Cepheid 
($\mu_\alpha^* = -6.15 \pm 2.8 \,\rm{mas\,yr}^{-1}$, $\mu_\delta = 12.89 \pm
2.9\, \rm{mas\,yr}^{-1}$, PPMXL). 

\paragraph{R\,Cru, T\,Cru and NGC\,4349}The two Cepheids R\,Cru and
T\,Cru have previously been considered as members of the open cluster NGC\,4349 
\citep{2002AJ....124.2931T}, although
they are no longer  listed in \cite{2010Ap&SS.326..219T}. Our results are
very clearly inconsistent with either Cepheid's membership in this cluster.
However, both Cepheids have very similar parallaxes and proper motions, and
lie close to the open cluster Loden\,624. Little information is available for
this cluster, and observational follow-up is warranted.

\paragraph{TV\,CMa and NGC\,2345} The membership
constraints parallax, proper motion (vanishes), and age, agree within their
respective uncertainties. The Cepheid's separation of $40$' from cluster center results in a
very low prior. Radial velocity differs by 
$\sim 20\,$km\,s$^{-1}$ between cluster and Cepheid, resulting in a low
likelihood. RV is the prime excluding constraint for this combo, and appears to
be robust.

\subsection{New Combos}\label{sec:newAppendix}
In the following subsections, we discuss possible membership for
selected Combos listed in Tab.\,\ref{tab:newCombos}. Observational follow-up is warranted
for all the inconclusive Combos in the table, even if they are not discussed 
here in detail. Further information can be found in the data compiled that is
available online, cf. Tab.\,\ref{tab:onlinetable}. We furthermore remark that the ASAS targets in
Tab.\,\ref{tab:newCombos} are particularly worthy of follow up, since they have
not yet received much attention.

\subsubsection{Inconclusive Combos}\label{sec:newInconclusive}

\paragraph{Y\,Car and ASCC\,60} Y\,Car is
a double-mode Cepheid in a triple system with a B9.0V companion \citep{1992ApJ...385..680E} on a known orbit (see
the \citealt{2003IBVS.5394....1S} binary Cepheids
database\footnote{\texttt{http://www.konkoly.hu/CEP/intro.html}}). The Cepheid
lies well inside ASCC\,60's projected core.
Since the RV of the cluster in K$05$ was measured on a single star and is
identical to Y\,Car's $v_\gamma$, we cannot consider this a valid membership
constraint.
Proper motion, on the other hand, is clearly measurable and consistent
with membership. Reddening is slightly larger (by 0.07\,mag) for the Cepheid
than for the cluster, and the absolute magnitude of Y\,Car estimated by
\cite{1992ApJ...385..680E} indicates a distance modulus incompatible with that
of the cluster, though we note that due to the sparsity of this cluster, a
revised distance estimate would be useful.
A detailed photometric and radial velocity study of ASCC\,$60$ is required to
conclude on the possible membership of Y\,Car. 
We furthermore note the presence of another Cepheid, CR\,Car inside
ASCC\,60's core radius. This Combo, however, is clearly inconsistent with
membership. CR\,Car lies in the background of the cluster.

\paragraph{VZ\,CMa and Ruprecht\,18} This combination yields a  
likelihood of $59\,\%$, since metallicity and age are in excellent
agreement between the cluster and the Cepheid.
Proper motion is better discernible for the Cepheid than for the cluster, it
seems, and color excess is $0.14$\,mag less strong for the Cepheid than
for the cluster. According to D02, the cluster (1.1\,kpc) lies bit
closer than the Cepheid ($1.3$\,kpc, from the Fernie database).  
In summary, both Cepheid and cluster require detailed follow-up for reddening
and parallax, before we can conclude on the question of membership.

\paragraph{WZ\,Car and ASCC\,63}
Likelihood and prior both tend to clearly exclude WZ\,Car's membership
in ASCC\,63; the large projected distance from ASCC\,63's core
(33\,pc assuming membership) lends further support to this interpretation.  
However, the likelihood computed is completely dominated by the extreme mismatch
in line-of-sight velocities ($\delta v_{\rm rad} = 120$\,km\,s$^{-1}$).
Looking at the other constraints, however, we find that age, reddening,
parallax (IRSB), and proper motion strongly suggest membership.
Suspiciously, the cluster RV is based on only 2 stars, and may therefore not be
reliable, or point towards an ejection event.
We therefore judge this Combo inconclusive and stress the need for
follow-up of the cluster.

\subsubsection{Unlikely Combos}\label{sec:newUnlikely}
\begin{table*}
 \centering 
 \caption{New Combos inconsistent with membership, despite high priors.
 Columns are described in Tab.\,\ref{tab:literature}.}
\begin{tabular}{@{}llccccccllll@{}} 
\hline 
Cluster 	 & Cepheid 	 & \multicolumn{6}{|c|}{Constraints} 	 & R$_{\rm cl}$ 	 & $P(A)$ 	 & $P(B \vert A)$ 	 & $P(A \vert B)$ \\
 & & $\varpi$ & $v_r$ & $\mu_\alpha^*$ & $\mu_\delta$ & [Fe/H] & age & [pc] &  & & \\
\hline 
Alessi 5	 & EY Car	 & $\circ$ & $\bullet$ & $\bullet$ & $\bullet$ & $\circ$ & $\circ$	 & 0.6	 & 1.0	 & 0.725	 & 0.725  \\  
Koposov 12	 & CO Aur	 & $\circ$ & $\circ$ & $2.0\sigma$ & $1.9\sigma$ & $\circ$ & $\circ$	 & 3.5	 & 1.0	 & 0.023	 & 0.023  \\  
Turner 1	 & S Vul	 & $2.8\sigma$ & $\circ$ & $\bullet$ & $1.6\sigma$ & $\circ$ & $1.3\sigma$	 & 0.2	 & 1.0$^{*}$	 & 0.015	 & 0.015  \\  
NGC 5045	 & NSV 19655	 & $\circ$ & $\circ$ & $2.2\sigma$ & $2.2\sigma$ & $\circ$ & $\circ$	 & 1.8	 & 1.0	 & 0.008	 & 0.008  \\  
Ruprecht 18	 & AO CMa	 & $3.8\sigma$ & $\circ$ & $\bullet$ & $1.4\sigma$ & $\bullet$ & $\bullet$	 & 0.5	 & 1.0	 & 0.005	 & 0.005  \\  
Collinder 173	 & AH Vel	 & $\circ$ & $\circ$ & $\circ$ & $\circ$ & $1.1\sigma$ & $2.3\sigma$	 & 11.8	 & 1.0$^{*}$	 & 0.036	 & 0.036  \\  
BH 34	 & ASAS J083130-4429.3	 & $3.9\sigma$ & $\circ$ & $1.6\sigma$ & $\bullet$ & $\circ$ & $2.7\sigma$	 & 0.8	 & 1.0$^{*}$	 & 0.0	 & 0.0  \\  
ASCC 60	 & CR Car	 & $4.4\sigma$ & $5.0\sigma$ & $1.3\sigma$ & $\bullet$ & $\circ$ & $3.2\sigma$	 & 0.7	 & 1.0	 & 0.0	 & 0.0  \\  
Collinder 173	 & ASAS J080101-4543.6	 & $\circ$ & $\circ$ & $\circ$ & $\circ$ & $\circ$ & $3.1\sigma$	 & 5.3	 & 1.0$^{*}$	 & 0.0	 & 0.0  \\  
Collinder 65	 & V1256 Tau	 & $4.7\sigma$ & $\circ$ & $\bullet$ & $1.3\sigma$ & $\circ$ & $4.7\sigma$	 & 3.4	 & 1.0	 & 0.0	 & 0.0  \\  
Melotte 25	 & NSVS 9444700	 & $\circ$ & $\circ$ & $\circ$ & $\circ$ & $\circ$ & $7.7\sigma$	 & 1.2	 & 1.0$^{*}$	 & 0.0	 & 0.0  \\  
Stock 2	 & GL Cas	 & $4.7\sigma$ & $\circ$ & $5.3\sigma$ & $7.6\sigma$ & $\bullet$ & $1.4\sigma$	 & 2.9	 & 1.0	 & 0.0	 & 0.0  \\  
Turner 11	 & X Cyg	 & $3.4\sigma$ & $\circ$ & $\circ$ & $\circ$ & $\circ$ & $5.2\sigma$	 & 0.0	 & 1.0$^{*}$	 & 0.0	 & 0.0  \\  
Turner 7	 & V Cen	 & $5.0\sigma$ & $\circ$ & $\circ$ & $\circ$ & $\circ$ & $15.5\sigma$	 & 0.0	 & 1.0$^{*}$	 & 0.0	 & 0.0  \\  
King 7	 & V933 Per	 & $\circ$ & $\circ$ & $\bullet$ & $2.0\sigma$ & $\circ$ & $5.1\sigma$	 & 3.4	 & 1.0	 & 0.0	 & 0.0  \\  
NGC 6639	 & X Sct	 & $2.9\sigma$ & $1.5\sigma$ & $\bullet$ & $1.7\sigma$ & $\circ$ & $5.0\sigma$	 & 1.2	 & 0.853	 & 0.0	 & 0.0  \\  
Teutsch 14a	 & ASAS J180342-2211.0	 & $1.8\sigma$ & $\circ$ & $\circ$ & $\circ$ & $\circ$ & $3.1\sigma$	 & 2.2	 & 0.814$^{*}$	 & 0.001	 & 0.001  \\  
NGC 6873	 & ASAS J200829+2105.5	 & $4.5\sigma$ & $\circ$ & $\circ$ & $\circ$ & $\circ$ & $\circ$	 & 3.8	 & 0.776$^{*}$	 & 0.0	 & 0.0  \\  
SAI 94	 & SX Vel	 & $3.8\sigma$ & $\circ$ & $\circ$ & $\circ$ & $\circ$ & $8.0\sigma$	 & 4.2	 & 0.761$^{*}$	 & 0.0	 & 0.0  \\  
Collinder 240	 & FR Car	 & $2.8\sigma$ & $\circ$ & $2.2\sigma$ & $\bullet$ & $\circ$ & $1.5\sigma$	 & 11.6	 & 0.733$^{*}$	 & 0.005	 & 0.004  \\  
NGC 6847	 & EZ Cyg	 & $2.4\sigma$ & $\circ$ & $1.2\sigma$ & $\bullet$ & $\circ$ & $5.4\sigma$	 & 8.7	 & 0.732$^{*}$	 & 0.0	 & 0.0  \\  
BH 23	 & AT Pup	 & $3.5\sigma$ & $2.6\sigma$ & $1.2\sigma$ & $\bullet$ & $\circ$ & $2.9\sigma$	 & 5.8	 & 0.616$^{*}$	 & 0.0	 & 0.0  \\  
Alessi-Teutsch 7	 & ASAS J082710-3825.9	 & $3.9\sigma$ & $\circ$ & $2.6\sigma$ & $\bullet$ & $\circ$ & $\bullet$	 & 17.8	 & 0.59$^{*}$	 & 0.0	 & 0.0  \\  
BH 164	 & AV Cir	 & $1.3\sigma$ & $1.3\sigma$ & $2.6\sigma$ & $5.0\sigma$ & $\circ$ & $\bullet$	 & 9.1	 & 0.572$^{*}$	 & 0.0	 & 0.0  \\  
Czernik 43	 & PW Cas	 & $2.8\sigma$ & $\circ$ & $\bullet$ & $\bullet$ & $\circ$ & $1.1\sigma$	 & 2.4	 & 0.565	 & 0.044	 & 0.025  \\  
\hline 
\end{tabular}
\label{tab:nonCombos}
\end{table*} 

\paragraph{Y\,Sgr and IC\,4725 (M25)} The parallaxes employed (Cepheid
$\varpi_{\rm{Cep}} = 2.13\pm 0.29$ from \citealt{2007AJ....133.1810B} and
Cluster $\varpi_{\rm{Cl}} = 1.61\pm 0.32$) in the calculation nearly agree
within their respective error budgets. However, color excess is 0.3\,mag
lower for the Cepheid, which indicates that Y\,Sgr lies in the foreground of
M\,25.

Using the well-established cluster member U\,Sgr as a point of
reference, we remark that proper motion, age, and metallicity are in excellent
agreement between both Cepheids. $v_\gamma$ differs slightly between the two,
which could be explained by the known binarity of U\,Sgr and Y\,Sgr. The only
significant discrepancy is in distance, which might be explained by
uncertainties in extinction, since M\,25 lies in the Orion arm (XHIP).
However, we calculate a very low prior for this Combo, and if Y\,Sgr
were a cluster member, it would lie at a large distance of $27$\,pc from cluster
center. From these considerations, it appears that Y\,Sgr is an unlikely
cluster member candidate.

\paragraph{BD+47\,4225 (GSC 03642-02459) and Aveni-Hunter 1}
The star was classified as a Cepheid based on HAT
data by \cite{2002PASP..114..974B} with a light curve that suggests
fundamental-mode pulsation. It lies barely outside $r_{\rm{c}}$ of cluster
Aveni-Hunter\,1. Unfortunately, proper motion and age are the only 
available membership constraints available in the literature compiled. 
Cluster and Cepheid appear to be co-moving in proper motion. Furthermore, the pulsational age of the 
Cepheid is spot-on with the cluster. However, a very rough distance estimate
using the V-band magnitude ($10.47$) found in the Guide
Star Catalog V.\,2.3.2 \citep{2008AJ....136..735L} yields a distance of
$2.5$\,kpc for the Cepheid, which is $5$ times the cluster distance. 
Since both objects are thus far not very well-studied, we highlight the need for
follow-up of both cluster and Cepheid.

\subsubsection{Non-members of interest}\label{sec:newNonMembers}

Non-member Combos of interest are listed in
Tab.\,\ref{tab:nonCombos}. We here present Combos with $P(A) = 1$, as well as
others with high priors and information on parallax. Some of the Cepheids
listed here belong to other open clusters, e.g. V\,Cen or X\,Cyg, or OB
associations, e.g. S\,Vul (T10). Below, we discuss one of these cases, EY\,Car,
since the membership probability computed is high.

\paragraph{EY\,Car and Alessi\,5} The beat Cepheid EY\,Car lies within
the core radius of Alessi\,5. The available (kinematic only) membership
constraints are consistent with, but not very close to, each other. However, the
literature distance from the Fernie database places the Cepheid nearly 2\,kpc
farther than the cluster, and the larger magnitude of proper motion is
consistent with a foreground cluster. EY\,Car is thus not a cluster member, 
mentioned here only due to the high membership probability computed.

\begin{table*}
\begin{tabular}{@{}llrcccrrrrrr@{}}
\hline
Cluster & Cepheid & R & P(A) & $P(B|A)$ & $P(A|B)$ & $\Delta\varpi$ & $\Delta v_r$ & $\Delta \mu_\alpha^*$ & $\Delta \mu_\delta$ & $\Delta \rm{[Fe/H]}$ & $\Delta \log{a}$ \\
 & & & & & & $\sigma_{\varpi}$ & $\sigma_{v_r}$ & $\sigma_{\mu_\alpha^*}$ & $\sigma_{\mu_\delta}$ & $\sigma_{\rm{[Fe/H]}}$ & $\sigma_{\log{a}}$ \\ 
 & & RefR & & &  & Refs$_{\varpi}$ & Refs$_{v_r}$ & Refs$_{\mu_\alpha^*}$ & Refs$_{\mu_\delta}$ & Refs$_{\rm{[Fe/H]}}$ & Refs$_{\log{a}}$ \\ 
\hline

IC 4725 & U Sgr &    1.570 &  1.00000 &  0.98373 &  0.98373 &   -0.114 &   -0.090 &   -0.850 &    0.780 &    0.040 &    0.210 \\ 
 &  &  &  &  &  &    0.328 &    3.606 &    2.983 &    2.891 &    0.162 &    0.250 \\ 
 &  & K05 &  &  &  & d,s,d,k,ks,V & MMB,* & t,h & d,ks & & d,FU \vspace{0.1cm} \\ 
NGC 7790 & CF Cas &    1.054 &  0.95454 &  0.97532 &  0.93098 &    0.056 &    0.399 &   -0.430 &   -0.090 &  &   -0.100 \\ 
 &  &  &  &  &  &    0.070 &    3.606 &    3.179 &    2.410 &  &    0.263 \\ 
 &  & B11 &  &  &  & d,p,d,k,ks,G & MMU,* & BDW,h & & & d,FU \vspace{0.1cm} \\ 
NGC 129 & DL Cas &    0.421 &  1.00000 &  0.85700 &  0.85700 &    0.063 &   -0.880 &    1.360 &    2.810 &  &    0.181 \\ 
 &  &  &  &  &  &    0.127 &    3.606 &    3.020 &    2.921 &  &    0.255 \\ 
 &  & K12 &  &  &  & d,p,d,k,ks,G & MMU,* & t,h & & & d,FU \vspace{0.1cm} \\ 
Turner 9 & SU Cyg &    0.067 &  1.00000 &  0.80743 &  0.80743 &    0.074 &    5.791 &    0.370 &    0.260 &  &    0.282 \\ 
 &  &  &  &  &  &    0.250 &    6.888 &    1.916 &    2.108 &  &    0.234 \\ 
 &  & K05 &  &  &  & d,s,d,k,ks,G & K07,* & K05,h & & & d,FU  \\ 
NGC 1647 & SZ Tau &  127.84 &  0.00000 &  0.04691 &  0.00000 &    0.060 &   -6.499 &    0.200 &    3.910 &  &    0.272 \\ 
 &  &  &  &  &  &    0.376 &    3.606 &    1.630 &    1.527 &  &    0.231 \\ 
 &  & K05 &  &  &  & d,s,d,k,ks,G & MMU,* & t,h & & & d,FO \vspace{0.1cm} \\ 
NGC 2345 & TV CMa &   38.208 &  0.00069 &  0.00001 &  0.00000 &    0.043 &   20.190 &   -0.720 &    1.290 &  &   -0.009 \\ 
 &  &  &  &  &  &    0.092 &    3.606 &    1.836 &    2.208 &  &    0.256 \\ 
 &  & B11 &  &  &  & d,p,d,f,f,A & MMU,f & KHA,h & & & d,FU \vspace{0.1cm} \\ 
NGC 4349 & R Cru &   14.985 &  0.04771 &  0.00000 &  0.00000 &   -0.717 &    1.308 &    4.790 &    2.510 &   -0.250 &    0.518 \\ 
 &  &  &  &  &  &    0.106 &    3.606 &    2.280 &    2.229 &    0.117 &    0.228 \\ 
 &  & K05 &  &  &  & d,p,d,f,ks,b & MMU,* & BDW,h & d,l & & d,FU \vspace{0.1cm} \\ 
ASCC 61 & SX Car &   41.253 &  0.00056 &  0.91917 &  0.00051 &    0.042 &  &    2.060 &    0.950 &  &    0.110 \\ 
 &  &  &  &  &  &    0.122 &  &    2.985 &    2.480 &  &    0.250 \\ 
 &  & K05 &  &  &  & d,p,d,f,ks,A & & K05,h & & & d,FU \vspace{0.1cm} \\ 
ASCC 69 & S Mus &   39.173 &  0.00446 &  0.87929 &  0.00392 &   -0.166 &    3.723 &    0.270 &   -0.040 &  &    0.260 \\ 
 &  &  &  &  &  &    0.215 &   12.266 &    1.414 &    1.807 &  &    0.253 \\ 
 &  & K05 &  &  &  & d,s,d,fo,ks,** & K05,* & K05,h & & & d,FU \vspace{0.1cm} \\ 
NGC 129 & V379 Cas &   42.900 &  0.00000 &  0.89594 &  0.00000 &  &    0.440 &    2.960 &    1.320 &  &    0.053 \\ 
 &  &  &  &  &  &  &    3.606 &    3.194 &    3.194 &  &    0.251 \\ 
 &  & K12 &  &  &  & & MMU,f & t,P & & & d,FO \vspace{0.1cm} \\ 
ASCC 60 & Y Car &    1.228 &  1.00000 &  0.78642 &  0.78642 &  &   -2.100 &   -1.520 &   -0.850 &  &  \\ 
 &  &  &  &  &  &  &    7.985 &    2.033 &    1.291 &  &  \\ 
 &  & K05 &  &  &  &  & K07,* & K05,h & & \vspace{0.1cm} \\ 
\multicolumn{12}{|c|}{... more data online ...} 	\vspace{0.1cm} \\
\hline
\end{tabular}
\caption{An excerpt of the data provided in the machine-readable online table. For each cluster-Cepheid combination we provide the parameters separation (R), prior, likelihood, and combined membership probability, as well as the differences between the individual membership constraints used (defined as cluster value minus Cepheid value), the combined error budgets adopted, and relevant references. For RefR, the cluster radius reference is provided. In column Refs$_{\varpi}$, references are listed in the following order: $\varpi_{\rm{Cl}}$, $\varpi_{\rm{Cep}}$, E(B-V)$_{\rm{Cl}}$, E(B-V)$_{\rm{Cep}}$, Cepheid $\langle m_{\rm{V}} \rangle$, Cepheid period (the latter two are relevant if a PLR-based distance was used). For the remaining columns, two references are given; the first corresponds to the cluster, the second to the Cepheid.
Due to spatial constraints, we abbreviate references in the following way: 
`a' - values based on ASAS photometry, 
`b' - values based on Berdnikov photometry, 
`d' - \citet{Dias} catalog, 
`f' - Fernie database, 
`fo' - \citet{2007A&A...476...73F},
`g' - GCVS, 
`h' - \citep{2007ASSL..350.....V}, 
`k' - \citet{2008MNRAS.389.1336K}, 
`ks' - \citet{2009A&A...504..959K},
`l' - \citet{2011AJ....142..136L}, 
`p' - parallaxes computed from PLR-based distances, 
`P' - PPMXL \citep{2010AJ....139.2440R}, 
`s' - \citet{2011A&A...534A..94S}, 
`t' - \citet{2001A&A...376..441D,2002A&A...388..168D},
`v' - VSX, 
`*' - newly determined $v_\gamma$ used, 
`**' - period improved using RV data,
`B11' - \citet{Buko11},
`FO' - first overtone ages from \citet{2005ApJ...621..966B},
`FU' - fundamental mode ages from \citet{2005ApJ...621..966B},
`K05' - \citet{2005A&A...438.1163K,2005A&A...440..403K},
`K07' - \citet{2007AN....328..889K},
`K12' - \citet{2012A&A...543A.156K},
`MMB, MMU, BDW, KHA' - see the references list in \citet{Dias}. 
The complete list of 3974 combinations can be retrieved from the online appendix to the paper.}
\label{tab:onlinetable}
\end{table*}

\label{lastpage}

\end{document}